\documentclass[11pt]{article}
\input{psfig}
\newcommand{\Se}[1]{\section{#1}\label{#1}}
\newcommand{\Sse}[1]{\subsection{#1}\label{#1}}
\newcommand{\Ssse}[1]{\subsubsection{#1}\label{#1}}

\newcommand{\Bi}{\begin{itemize}}
\newcommand{\Ei}{\end{itemize}}
\newcommand{\Bn}{\begin{enumerate}}
\newcommand{\En}{\end{enumerate}}
\newcommand{\Bd}{\begin{description}}
\newcommand{\Ed}{\end{description}}
\newcommand{\Bq}[1]{\begin{equation}\label{#1}}
\newcommand{\Eq}{\end{equation}}
\newcommand{\Bqn}[1]{\begin{eqnarray}\label{#1}}
\newcommand{\Eqn}{\end{eqnarray}}
\newcommand{\BTab} { \vspace{0.3cm} \begin{table}[phtb] \centering }
\newcommand{\ETab}[2] {\caption{#2} \label{#1:tab} \end{table} \vspace{0.3cm} }
\newcommand{\BTAB} { \vspace{0.3cm} \begin{table*}[phtb] \centering }
\newcommand{\ETAB}[2] {\caption{#2} \label{#1:tab} \end{table*} \vspace{0.3cm} }

\newcommand{\junk}[1] {}
\newif\ifsizedfigures \sizedfiguresfalse
\newif\ifincludefigures \includefiguresfalse
\newif\iffigurenamesinmargin \figurenamesinmarginfalse
\newif\ifpdf\ifx\pdfoutput\undefined\pdffalse\else\pdfoutput=1\pdftrue\fi

\usepackage{times}
\usepackage[pdftex]{graphicx}
\newcommand{\mycomment}[1]{}

\newcommand{\figw}[3]
{
\begin{figure}
\centerline{\psfig{figure=fig/#1.eps,width=#2}}
\caption{\label{#1:fig} \small #3}
\end{figure}
}

\begin{document}
\title{\bf Analysis of 802.11b MAC: A QoS, Fairness, and Performance Perspective
}

\author{Srikant Sharma\\
        Department of Computer Science\\
        Stony Brook University\\
	Stony Brook, NY 11794-4400\\
        {srikant@cs.sunysb.edu}\\
}
\date{}
\maketitle

\begin{abstract}

Wireless LANs have achieved a tremendous amount of growth in recent
years. Among various wireless LAN technologies, the IEEE 802.11b based
wireless LAN technology can be cited as the most prominent technology
today. Despite being widely deployed, 802.11b cannot be termed as a well
matured technology. Although 802.11b is adequate for basic connectivity
and packet switching, It is evident that there is ample scope for its
improvement in areas like quality of service, fairness, performance,
security, etc. In this survey report, we identify and argue that the
Medium Access Controller for 802.11b networks is the prime area for these
improvements. To enunciate our claims we highlight some of the quality
of service, fairness, and performance issues related to 802.11b MAC.
We also describe and analyze some of the current research aimed at
addressing these issues.  We then propose a novel scheme called the {\em
Intelligent Collision Avoidance}, seeking to enhance the MAC to address
some of the performance issues in 802.11b and similar networks.

\end{abstract}

\Se{Introduction}


Although the concept of wireless LANs has existed since late 1970s,
the WLAN technology started gaining its momentum only in late 1990s
and today it has become a ubiquitous networking technology. The reason
behind the recent explosive growth of this technology can be attributed
to multiple factors, such as, technological advances in error correcting
codes, modulation techniques, processing power on network interfaces,
availability of unlicensed radio spectrum, and most importantly, the
need for some kind of tetherless connectivity and mobility.

Today there exist multiple wireless LAN technologies, such as, Wi-Fi,
Bluetooth, HiperLAN, HomeRF, etc. All of these technologies operate in
the 2.4GHz ISM (Industrial, Scientific, and Medical) radio spectrum. Each
technology has its own niche depending on the deployment requirements of
the wireless LANs. Bluetooth is mainly used as a cable replacement RF
technology for short range communications. It is used to interconnect
portable devices, such as, cellular phones, laptops, palmtops, etc.,
without the need to carry interconnecting cables. It is capable of
providing data rates upto 700 Kbps and supports upto three voice channels
at 64 Kbps. HomeRF is used for wireless home networking for devices like,
intelligent home appliances, laptops, smart pads etc. It has a range of
upto 50 meters which is adequate for short scale networks. It supports
data rates upto 1.6 Mbps which is low by todays WLAN norms. HiperLAN
is a family of four different wireless technologies classified as types
1-4. HiperLAN-1 operates in 5GHz radio spectrum is capable of supporting
upto 23 Mbps at a range of 50 meters.  One of the prime features of
HiperLAN is its support for Wireless ATM.  WATM is the extension of
ATM capabilities, such as QoS features, etc., to wireless networks.
{Wi-Fi} technology is based on IEEE 802.11b~\cite{80211b} standard. It
operates in the unlicensed 2.4 GHz radio spectrum, uses direct-sequence
spread-spectrum (DSSS) for modulation, supports variable data rates upto
11 Mbps, and has a range of about 50 meters.


Out of all these technologies, IEEE 802.11b or Wi-Fi is the technology
which has received the widest market acceptance.  The popularity of this
standard is aptly reflected in portable computer vendors' decision to
integrate 802.11b wireless network adapters with notebook computers. A
market forecast by the Gartner \cite{gartner} group predicts that by the
end of year 2005 almost 95\% of notebook computers will be equipped with
802.11b cards. Further, by the end of year 2002 the 802.11b penetration
in corporate LANs is expected to reach up to 50\%, from the current
level of around 20\%.  Almost all PDA vendors are starting to support
the 802.11b technology in the newer-generation PDAs on the market. The
widespread availability of 802.11b on wireless devices coupled with
continuous cost reduction is also a strong indication of exponential
growth of the 802.11b technology.


IEEE 802.11b LANs can be deployed in either {\em ad hoc} configuration
or {\em infrastructure} configuration. The ad hoc configuration refers
to the peer-to-peer setup where a bunch of devices with 802.11b network
interface cards (NICs) can establish a network and communicate with
each other without any infrastructural support.  The connectivity of
the nodes in this network is limited to their peers. On the other hand,
the infrastructure or the access-point setup uses a central access-point
(base-station) to form a network. The access-point is usually connected
to a wired network as a bridge for next hop connectivity. Every packet
transmitted by a wireless node is destined for the access-point which
takes care of further routing/switching.

Most of the corporate and large scale wireless networks are setup in
the infrastructure mode of operation.  There are two different classes
of infrastructure operation. These are {\em basic service set} (BSS)
and {\em extended services set} (ESS). In BSS configuration each wireless
node is {\em associated} with an access-point and this association remains
unchanged indefinitely, whereas, in ESS a mobile node can roam around
and {\em disassociate} from current access-point and associate with a new
access-point or {\em re-associate} with the previous access-points. The
ESS is basically meant to provide roaming support.


IEEE 802.11b technology has achieved a huge level of penetration in the
wireless networking arena.  It is being regarded as the de facto wireless
standard for wireless LANs. Although the dependence on 802.11b is growing,
it cannot be termed as a very well matured wireless LAN technology. The
technology, though adequate for basic connectivity and packet switching,
falls short of expectations when it comes to issues like, quality of
service, fairness, performance, security, etc. The wireless research
community is persistently finding different ways to improve this
technology and bridge the shortcomings. The channel access protocol
(MAC) used by 802.11b networks is known to have several performance
related issues. The quality of service for applications using these
networks is practically non existent. The protocol is known to exhibit
unfairness for different streams in terms of channel allocation. The
security aspect of these networks is known to have several problems.

This survey attempts to highlight the QoS, fairness, and performance
issues in 802.11b networks and various research related to MAC
enhancements to address these issues.  The report is organized
as follows.  In Section~\ref{IEEE 802.11b MAC overview}, we give an
overview of 802.11b channel access protocol.  In Section~\ref{Quality
of Service}, we briefly look at the QoS efforts for 802.11b networks.
We examine the fairness problem of 802.11b MAC in Section~\ref{Fairness}.
In Section~\ref{Performance}, we look at current performance related
research for 802.11b MAC.  In Section~\ref{Intelligent Collision
Avoidance} we propose and analyze enhancements to 802.11b MAC protocol
to improve throughput performance at the expense of power consumption.
Finally, in Section~\ref{Conclusions} we present our conclusions and
discuss the work that we are planning to pursue in future.

\Se{IEEE 802.11b MAC overview}

IEEE 802.11b is a standard for Medium Access Control (MAC) and Physical
Layer (PHY) specifications for wireless LANs. The PHY specifications
deal with modulations techniques, error correcting codes, radio
characteristics, physical layer convergence, and other signaling related
issues.

IEEE 802.11b MAC protocol is based on the CSMA/CA~\cite{maca} protocol
which uses physical carrier sense as well as virtual carrier sense to
avoid collisions and packet loss. Physical carrier sense is used to
avoid collisions at the sender, whereas, virtual carrier sense is used
to avoid collisions at the receiver and address the {\em hidden node}
problem present in wireless networks. The virtual carrier sense uses
regular Request To Send (RTS) and Clear To Send (CTS) channel reservation
mechanism.  802.11b MAC improves the link layer reliability by including
explicit ACKs for each data frame. Upon failure to receive an ACK,
the data frame is repeatedly retransmitted till an ACK is received. The
maximum number of retransmissions is a configurable parameter for each
individual node and is usually set to seven.  Thus each successful
transmission follows the so-called {\em 4-way handshake} protocol of
RTS-CTS-DATA-ACK.  A node may choose to disable the virtual carrier sense
to reduce its overhead when the probability of existence of hidden nodes
is known to be small.

802.11b MAC includes two coordination functions for channel access,
namely, Distributed Coordination Function (DCF) and Point Coordination
Function (PCF). The DCF specifies channel contention mechanism for
normal mode of operation, whereas, PCF specifies a mechanism for channel
access in a contention free fashion.  PCF requires the presence of a
{\em point coordinator} (PC) and can be used only in infrastructure
mode of operation. We will be describing the details of DCF in the
following part of this section. We will be looking at the details of
PCF in Section~\ref{Quality of Service}.

\Sse{Distributed Coordination Function (DCF)} 

In the normal mode of operation, the IEEE 802.11b MAC uses a Distributed
Coordination Function (DCF) for media access. DCF is an implementation
of CSMA/CA protocol which follows the 4-way handshaking protocol for
data transmissions.  In DCF, whenever a node is ready to transmit data,
it senses the channel to be idle for a period of Distributed Inter
Frame Spacing (DIFS). Following this, it generates a random backoff
timer chosen uniformly from the a [0, $\omega$-1] where $\omega$ is
the {\em contention window}.  Initially the contention window is set
to {$CW_{min}$} (16 for 802.11b). After the backoff timer expires,
the node sends a short Request To Send (RTS) message to the intended
receiver of data.  If this message is received properly by the receiver
and if it is able to receive any transmission, it responds back with a
short Clear To Send (CTS) message. A node may not be able to receive any
transmission if some other node in its vicinity has already reserved
the channel for packet reception or transmission. Both RTS and CTS
messages carry the duration information for which the channel is going
to be occupied by the proposed data transmission. Upon hearing RTS and
CTS, all other nodes in the vicinity of the sender and receiver update
their Network Allocation Vectors (NAVs)  with the information about the
duration for which the channel is going to be busy. NAV is essentially
a channel reservation vector. Thus, all nodes in the vicinity of the
sender and receiver defer their transmissions and receptions to avoid
collisions. The CTS message is followed by the DATA transmission which
is acknowledged by the receiver by sending an ACK message if the DATA
is received successfully. The data is repeatedly retransmitted in the
absence of ACKs till a threshold number of retransmissions are carried
out. Once the retransmissions exceed the threshold, the transmission is
assumed to be unsuccessful. After an unsuccessful transmission attempt,
the sender follows a binary exponential backoff (BEB) and doubles its
contention window size. This is done in order to reduce the channel
contention between nodes. The contention window is not incremented
further if it already equals {$CW_{max}$} (256 for 802.11b). After
every successful transmission, the contention window is reset back to
{$CW_{min}$}. RTS,CTS,DATA, and ACK are separated by a time spacing of
Short Inter Frame Space (SIFS). A timeline for DCF message exchanges is
shown in Figure~\ref{dcf:fig}. The sense period of DIFS is always larger
than SIFS. This ensures that no new transmission attempts interfere with
the ongoing transmission.

DCF provides a mechanism for collision avoidance by performing a virtual
carrier sense through RTS-CTS message exchanges. This is necessary to
solve the {\em hidden node problem}. However, a node may chose to resort
to a 2-way handshaking mechanism where data packets are transmitted
without RTS-CTS message exchanges. The packet is acknowledged back by
the receiver by responding with an ACK message. This mechanism can be
used for small packets where the overhead of RTS/CTS message exchange
can be traded off for small probability of collisions.

\figw{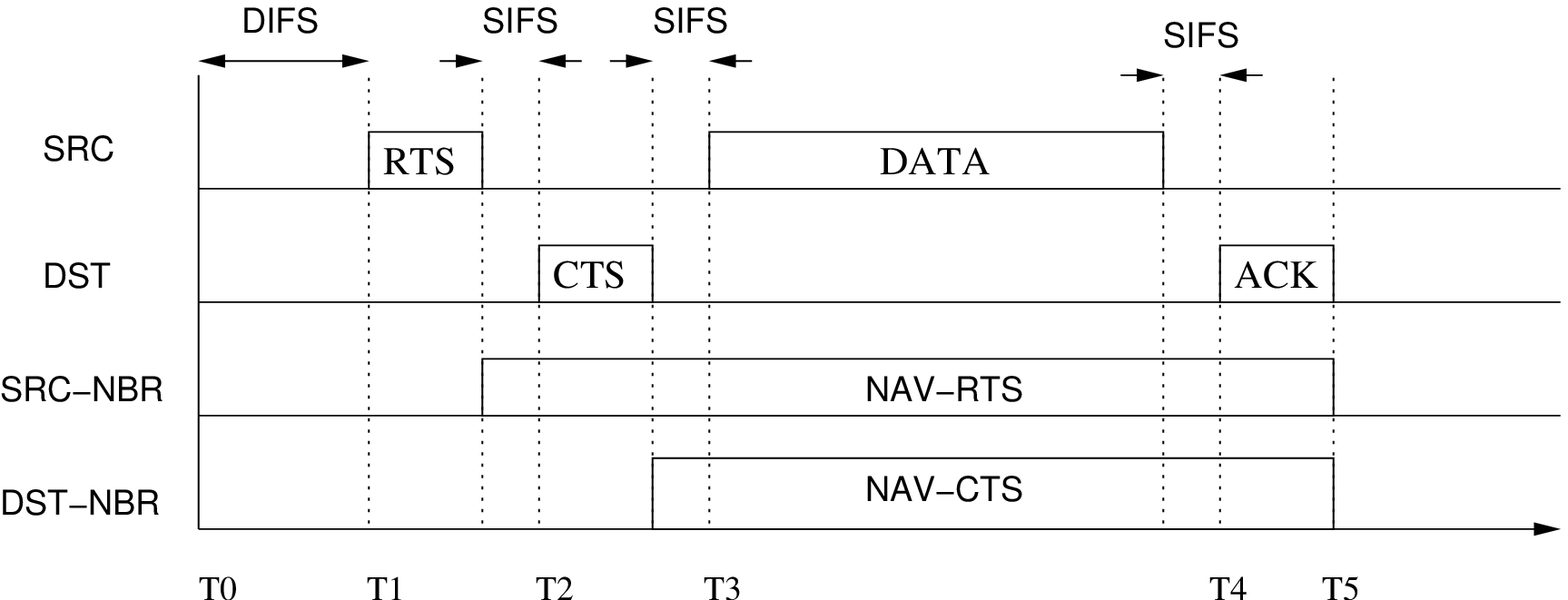}{5.5in}{\sl Message exchanges for DCF. Each DATA packet is
preceded by RTS and CTS messages. On hearing RTS, the nodes in the
vicinity of sender set their NAVs to the duration mentioned in RTS. On
hearing CTS, the nodes in the vicinity of receiver set their NAVs to
the duration mentioned in CTS. This causes in establishment of channel
reservation till the time the ACK is sent back to the sender.}

\Sse{Link Management}

Apart from channel access mechanism, 802.11b also provides additional
facilities such as power save features, security etc. Moreover,
802.11b MAC is a distributed entity where decisions regarding various
activities are taken independently at various nodes.  To distribute MAC
level information among the constituent nodes, 802.11b defines several
messages which are transmitted as link level frames.  The frames that are
exchanged between different nodes of an 802.11b network are classified
into three categories as follows:

\begin{itemize}
\item {\bf Management}
	\begin{itemize}
	\item{\{association, re-association, probe\}\{request, response\},
	and disassociation: These frames are used for affiliation activity
	of any node to a particular cell and access-point.}

	\item{authentication and de-authentication: These frames are
	used for security purposes.}

	\item{beacon: This frame is usually sent by an access-point in
	infrastructure mode.  In ad hoc mode, the first node to initiate
	an ad hoc network sends the beacon.  This frame carries important
	management information like time-stamp, supported rates, traffic
	indication maps etc.}

	\item{ATIM: Announcement Traffic Indication Message is a frame
	sent after every beacon frame. This is used by nodes utilizing
	the power save features of 802.11b to indicate their traffic
	pattern.} 
	\end{itemize}

\item{\bf Control}\\
	The frames like RTS, CTS, ACK, PS-Poll (power save poll) CF-Poll,
	CF-ACK, CF-END (Contention free channel information in PCF)
	are categorized as control frames.

\item{\bf Data}\\
	The data frames can be piggybacked with frames like CF-ACK,
	CF-Poll etc.

\end{itemize}

IEEE 802.11b also specifies a {\em fragmentation} mechanism using which
an 802.11b node can divide data packets into smaller frames and transmit
large packets when the frame error rates are high. The fragmentation
threshold can be a configurable value or can be dynamically determined
depending on the channel conditions.

To summarize, the 802.11b MAC layer provides, (1) Channel access
mechanism, (2) Link management facility, (3) Rudimentary security,
(4) Power management, and (5) Fragmentation.

\figw{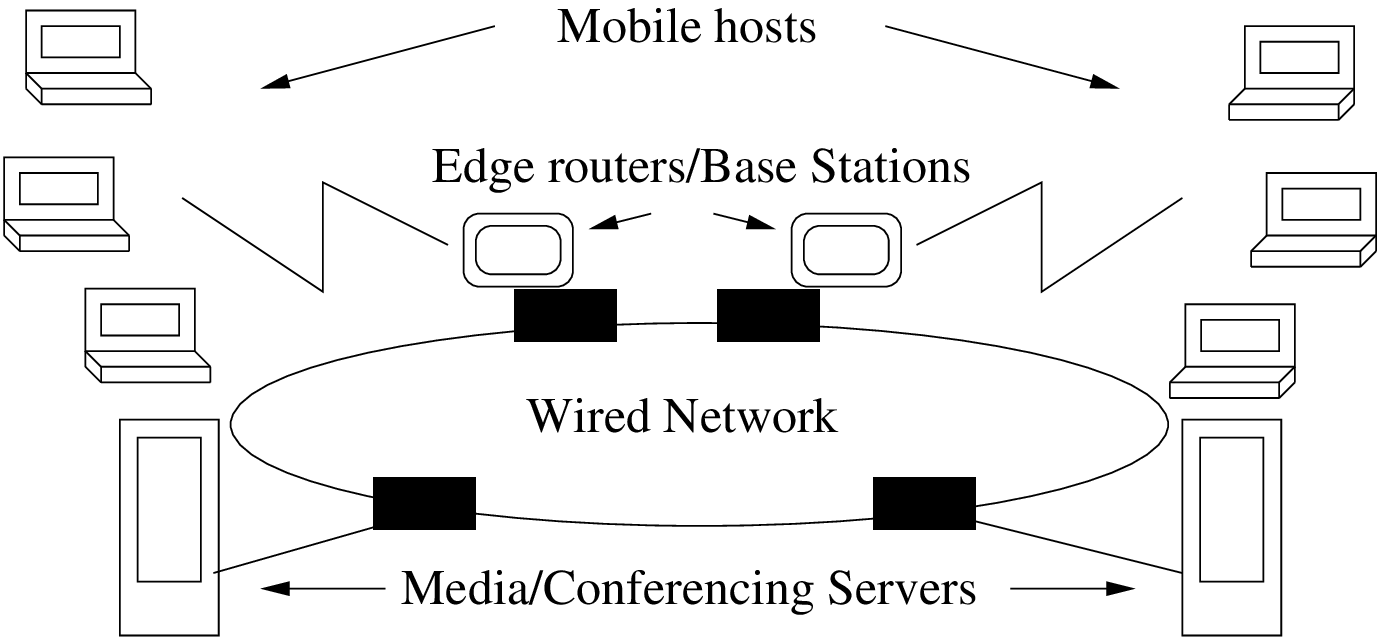}{4in}{\sl IEEE 802.11b LANs are usually at the periphery of
wired networks serving as the access networks. The servers are inside
the wired domain and clients are in wireless domain exploiting the
mobility advantage.}

\Se{Quality of Service}

IEEE 802.11b WLANs provide {\em best effort service} similar to their
wired counterpart Ethernet networks. Best effort service essentially
indicates that {\em every} data packet handed over to the 802.11b
interfaces receives similar treatment as other packets in terms of
delivery guarantees. Thus, from an application perspective, it receives
no Quality of Service guarantees from the network in terms of available
bandwidth, latency, jitters etc.  Some applications like media streaming
and conferencing are sensitive to packet latency and effective bandwidth
characteristics of the underlying network.  With the growing thrust on
use of wireless networks for media streaming and conferencing, it becomes
essential to provide a service that is not merely best effort service but
a service that is deterministic at certain level.  IEEE 802.11b WLANs are
usually deployed as {\em access networks} to the wired infrastructures for
mobile terminals as shown in Figure~\ref{network:fig}.  In wireless LAN
scenario, usually the media and conferencing servers reside in the faster
and reliable wired domain, whereas, the clients reside in wireless domain
exploiting the mobility features. Thus the importance of QoS is relatively
higher in the infrastructure setup rather than the peer-to-peer setup.

On the wired network, IETF's Integrated Services \cite{rsvp} and
Differentiated Services \cite{diff,tos} architectures are available to
support guaranteed QoS and traffic prioritization respectively above the
link layer. IEEE's 802.1p is a layer 2 traffic prioritization standard
for switched Ethernet environments. However, there are very limited QoS
solutions that exist for wireless LANs, particularly 802.11b networks.

\Sse{Point Coordination Function}

IEEE 802.11b standard provides a very rudimentary support for quality
of service in its infrastructure mode of operation. This support is
provided by the MAC layer in terms of {\em Point Coordinated Function}
(PCF). PCF is a MAC coordination facility that may exist on access-points
to differentiate between the traffic flows from different nodes.  PCF is
an optional capability for access points and its implementation is not
mandatory. Very few commercially available access-points for 802.11b
networks actually provide this facility. Moreover, there are no clear
mechanisms for individual nodes to participate in PCF and exploit the
quality of service mechanism provided by it.

PCF provides a very rudimentary form of QoS by allowing nodes to transmit
frames in a contention free manner. The access-point of a cell acts as a
coordinator called the {\em point coordinator} (PC) for that cell. All
nodes in 802.11b network obey the medium access rules of the PCF,
since these are based on DCF which is followed by all nodes.  The PC
grants a contention free channel access to individual nodes by polling
them for transmissions. On being polled, a node transmits a single frame
destined for any node in the network. All nodes which need some quality
of service are termed as CF-Pollable nodes. A node becomes CF-Pollable by
indicating its interest in being polled during its association with an
AP (PC). This results in the node being included in a CF-Pollable list
maintained by the PC. A node can withdraw itself from polling process
by performing a re-association.

In infrastructure mode of 802.11b, the time is divided into periodic {\em
superframes} which start with the so-called {\em beacon frames}. A beacon
frame in 802.11b is a management frame sent by an access point to carry
out time synchronization and deliver protocol related information to all
nodes. Beacon frames are periodically sent by the access points regardless
of PCF functionality. Each superframe is divided into two units, namely,
{\em Contention Free Period} (CFP) and {\em Contention Period} (CP).
CFP is the period when contention free channel access is provided by
the PC to individual nodes.  CP is the period when all nodes contend
for the channel using DCF.  If the PCF functionality is not provided by
the access point then entire superframe is the contention period. The
extent of division of a superframe into CFP and CP is determined by the
PC which can be arbitrary, but it is mandatory to have a CP of a minimum
duration that allows at least one node to transmit one frame under DCF.

\figw{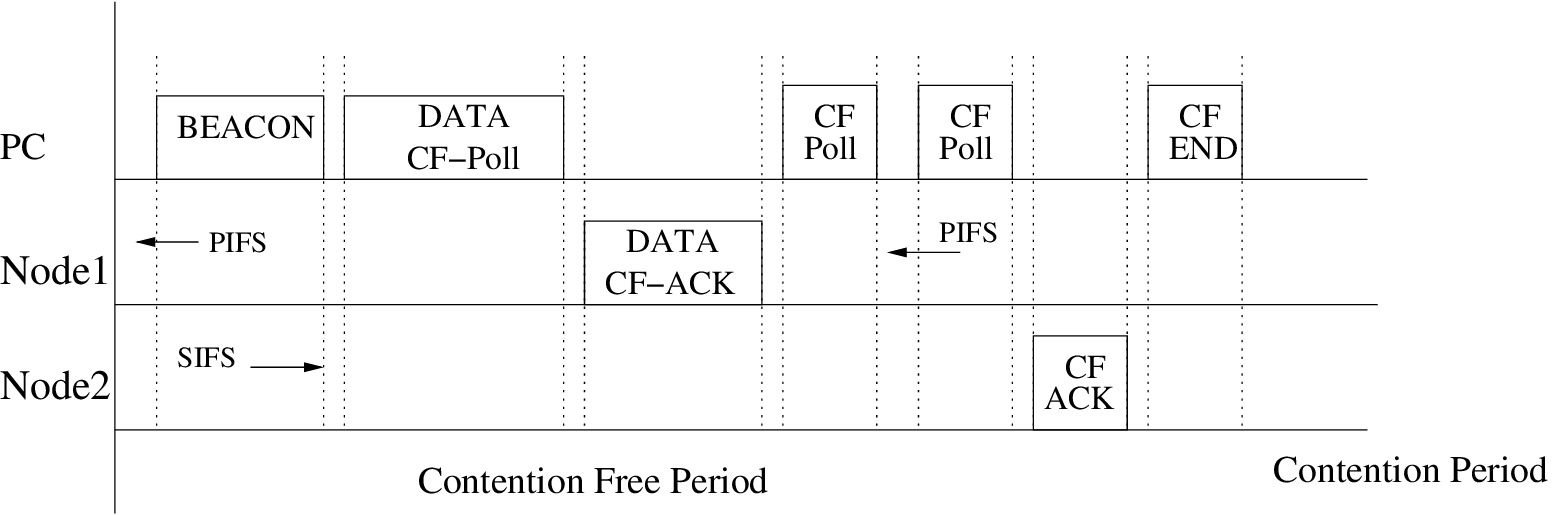}{6in}{\sl A superframe in 802.11b. Each superframe
is divided into contention free and contention periods. PC polls nodes
during CFP. If there is no response from a node, the PC polls next node
after PIFS. The nodes respond to polls with either DATA+ACK frames or
just ACK frames after SIFS interval. No nodes other than polled nodes
can access the channel since it is never idle for DIFS period during CFP.
The CFP ends when the PC sends CF-END frame. After this all nodes enter
CP where DCF is used to access the channel.}

Figure~\ref{superframe:fig} shows the activity of a wireless network
during a superframe. At the beginning of superframe the PC waits for
a period PCF Inter Frame Space (PIFS) and then transmits the beacon
frame. If the PC supports PCF and the list of nodes that are interested
in being polled is not empty, the PC sends a CF-Poll (or DATA+CF-Poll)
frame to one of the nodes after waiting for channel to be idle for SIFS.
In response, the node can respond with a DATA + CF-ACK or just CF-ACK
if no data is ready to be sent. The response is sent after sensing
the channel to be idle for an SIFS period.  If there is no response to
CF-Poll frame, the PC sends CP-Poll to next node after waiting for an
idle period of PIFS. At the end of CFP, the PC sends a CF-END frame
to begin the contention period using DCF. Thus in CFP, each polled
node transmits frames in a contention free manner. In CFP, RTS/CTS
handshaking is not carried out.  During the entire CFP the PC is in
control because it accesses channel after sensing the channel to be idle
for PIFS duration. PIFS is much smaller than DIFS which is the period for
which  every nodes in DCF should sense the channel to be idle. The shorter
duration of PIFS compared to DIFS ensures that no node can contend for the
channel except either the PC or the node that has been recently polled.

There are several problems with PCF that make it less attractive for
QoS utilization.  First and foremost, there is no guarantee of bandwidth
or any other QoS {\em parameters} except a contention free transmission
of a single frame. The channel share available to individual nodes
cannot be specified and it decreases with an increase in the number of
pollable nodes in the network. Second, there is no definite bound on next
poll being received by a node. The polls received from PC for a node
may not be periodic.  For flows with tight delay bounds and periodic
traffic, this may be unacceptable. Third, there is no guarantee that
the superframe itself is periodic. This is because the PC must sense the
channel to be idle for PIFS duration before sending the next beacon. If
the transmission of last frame from the previous superframe is prolonged
then the beacon transmission may not be strictly periodic. Fourth, the
size, rate and transmission time occupied by each frame is not constant.
Some frames may be fragmented and may be arbitrarily long (upto 2312
bytes) and hence may take longer than usual transmission time. This
will affect subsequent nodes. Some of the nodes may not be polled in a
particular superframe. In short, there is no clear admission control and
usage policy. Last, there is no clear way of interfacing an application
with this mechanism.  This makes PCF {\em a facility with no use}.

\Sse{Rether for IEEE 802.11b}

If one were to revamp PCF and come up with a scheme that is tailored to
provides QoS to applications, following enhancement would be required.\\

\noindent
{\bf Interface:}  There must be some interface through which an application
can specify its bandwidth requirements to the underlying QoS mechanism.\\
{\bf Admission Control:} There must be some admission control criteria so that the
available meager resources, once granted to certain applications, are not hijacked back.\\
{\bf Periodicity:} The channel access must be provided in a fairly periodic fashion
for delay sensitive and periodic traffic.\\
{\bf Isolation:} Once an application (or its flow) is admitted, the service
provided to it must not be affected by quirkiness of other applications and nodes.\\
{\bf Bandwidth Guarantee:} The channel access granted should reflect into an appropriate
bandwidth guarantee over a long term duration.\\

Rether~\cite{srikant} is a QoS mechanism for 802.11b networks which
provides bandwidth guarantees to individual flows. Rether borrows many
salient features of PCF and is a software solution residing in the network
stack of individual nodes. Rether addresses all of the above issues. In
addition, it does not require any changes at the MAC layer and does not
use the PCF capability which may not be available with every access point.

Like CFP and CP in PCF, Rether also has a notion of Real-Time (RT)
and Non Real-Time (NRT) periods. Analogous to the superframe, Rether
has a concept of a periodic cycle which corresponds to one set of RT
and NRT period. Rether strives to guarantee a contention free channel
access to all nodes during both RT and NRT period. During RT period,
the channel access is granted to nodes with bandwidth reservations,
and the NRT period is used to grant channel access to all nodes in the
wireless network. Similar to the mandatory requirement of CP in PCF,
Rether also has a notion of RT limit of bandwidth reservation beyond
which no new flows are granted bandwidth guarantees. This is done to
avoid the starvation of flows with no bandwidth reservations. Rether
grants channel access to individual nodes by sending so-called {\em
tokens}. The individual nodes signify their end of transmission by
sending and explicit {\em acknowledgment}.  Rether grants contention
free channel access even in NRT period by circulating the token in a {\em
round robin} fashion among the nodes without any bandwidth reservation.
In order to avoid fairness issues, the round robin nature is persistent
across consecutive NRT periods.

\begin{figure*}
\begin{minipage}[t]{3.1in}
\centerline{\psfig{figure=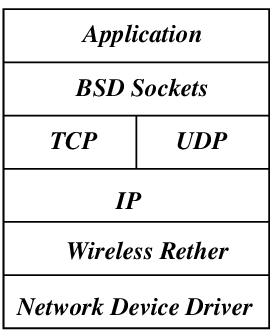,height=1.2in}}
\caption{\small{\sl Wireless Rether protocol layer. Wireless Rether is
implemented as a layer between the link-layer hardware's device driver and the
IP protocol layer. With this layering structure,  Wireless Rether is able to exercise
QoS-related control over all outgoing network connections.}}
\label{layer:fig}
\end{minipage}
\hspace{.1in}
\begin{minipage}[t]{3.1in}
\centerline{\psfig{figure=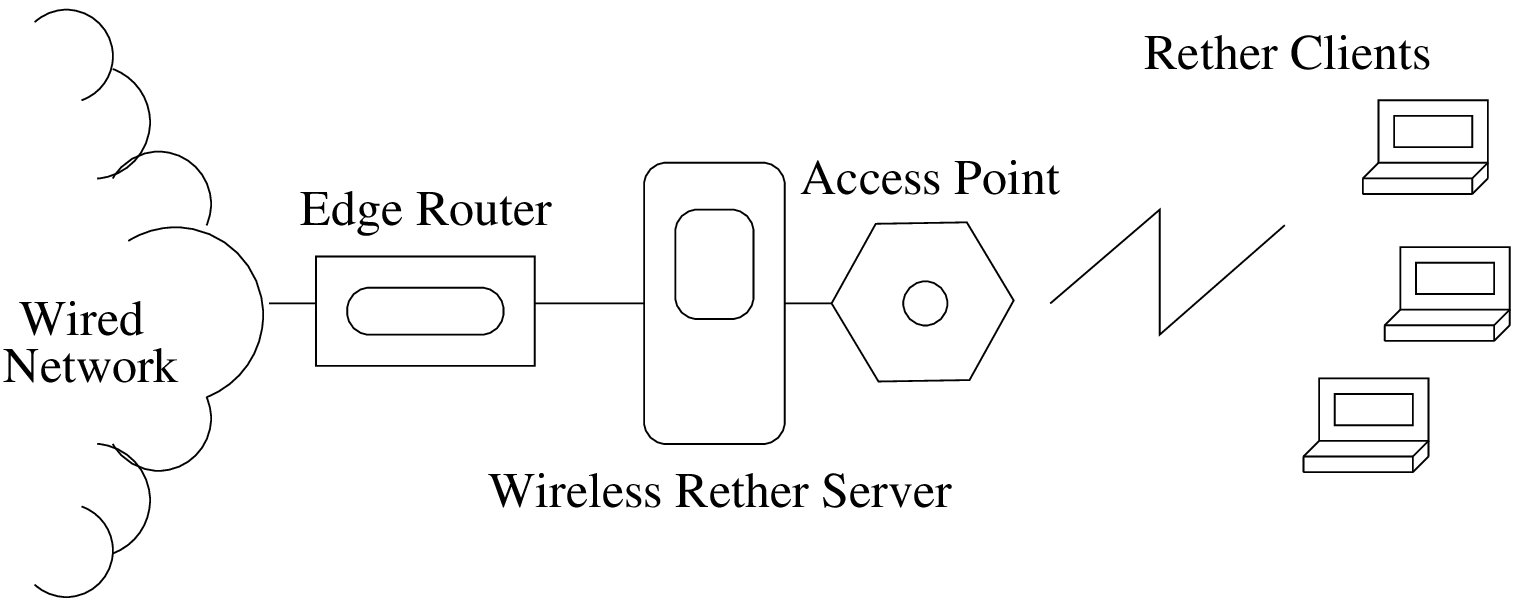,width=3in}}
\caption{\small{\sl Wireless Rether Architecture. Each edge network is
augmented with a Wireless Rether Server (WRS), which acts as a bridge between
the Layer-2 access point and the Layer-3 edge router. All wireless LAN
hosts are equipped with Wireless Rether client software.}}
\label{arch:fig}
\end{minipage}
\end{figure*}

As depicted in Figure~\ref{layer:fig}, Rether is implemented as a
software module which resides between the IP layer and the device
driver for the wireless interface.  It intercepts all the packets that
are handed over to the device driver and exercises control over their
transmission which follows established QoS policies. Rether follows
a client-server architecture. In an infrastructure wireless LAN, the
Wireless Rether Server (WRS) is co-located with the access point. All
nodes in the network understand the Rether protocol and are termed as
Wireless Rether Clients (WRC). The WRS is primarily responsible for
admission control and channel coordination.

Rether uses implicit bandwidth reservation mechanism based on port
signaling. The advantage of this mechanism comes from compatibility
with legacy applications which cannot be modified to carry out explicit
reservation requests. The reservation mapping is specified in system
policy files in terms of {\tt quintuples} like :

\centerline{\tt \{SrcAddr/Mask, DstAddr/Mask, SrcPrtRange, DstPrtRange, BW\}}

Rether module intercepts all outgoing packets on a node. Upon intercepting
the first packet corresponding to any flow, the WRC sends a reservation
request to the WRS if a corresponding policy specification exists. If
the request is accepted all packets corresponding to the flow are
maintained in a separate queue.  The WRS circulates the token among all
nodes who have established reservations. All WRCs dispatch packets from
their queues according to the bandwidth requirements upon reception
of tokens. The dispatching is limited to bandwidth share or allocated
slice of the cycle which ever occurs first. This limitation helps in
providing isolation between nodes with bad radio characteristics and
normal nodes. Once all nodes with reservations are served, WRS switches
to NRT mode and the clients without any reservation are provided with
tokens in a round robin fashion.

With this architecture, Rether provides an effective bandwidth guarantee
scheme which can be readily used by applications. The guarantees are
provided by granting a contention free channel access to all nodes
transmitting packets. Further, Rether is an all-software protocol which
does not need any modification to the underlying MAC layer.

\Sse{IEEE 802.11e}

The market thrust on wireless LAN technologies and the requirement of
QoS support in these networks has led to the establishment of an IEEE
standardization working group to enhance 802.11 to provide applications
with QoS support. Upcoming standard 802.11e tries to address the QoS
problems faced by wireless LANs based on 802.11 specifications.

The QoS support in 802.11e is provided in two forms. First, it supports
a priority based best-effort service similar to diffserv. Second, it
supports parameterized QoS for the benefit of applications requiring
QoS for different flows. 802.11e achieves this by enhancing the 802.11
DCF and PCF functionality, and by providing a signaling mechanism for
parameterized QoS.

The enhanced MAC protocols in 802.11e are EDCF and EPCF. Both EDCF and
EPCF are commonly refered as the Hybrid Coordinated Functions (HCF). The
priority based best effort service is provided by the EDCF. This is done
by introducing so-called {\em traffic categories}. Frames corresponding to
different traffic categories are now transmitted through different backoff
instances. Each traffic category has an associated independent backoff
instance. The scheduling of frames for every traffic category is done the
same way as in DCF. The differentiation in the priority is achieved by
setting different probabilities for different categories for winning the
channel contention.  The probability is changed by varying the so-called
Arbitration Inter Frame Space (AIFS) which is the listen interval
for channel contention. AIFS is analogous to the DIFS period in DCF.
For each traffic category, the value of AIFS determines the priority.
With lower AIFS values, the listen interval required for channel
contention is lower and hence the probability of winning the channel
contention is higher. For compatibility with legacy DCF, AIFS should be
at least equal to DIFS.  The backoff procedure in case of collisions is
similar to that in DCF. The main distinguishing factor is the method by
which the contention window is expanded after collisions. In DCF, the
contention window is always doubled.  Whereas in EDCF, the contention
window is expanded by a predetermined {\em persistence factor} (PF).
For example, in DCF scenario the PF is always 2 since the contention
window is always doubled.

A single node can have upto eight traffic categories. These different
categories are realized as eight different virtual nodes with varying
parameters, such as, AIFS, CW, and PF. These parameters are responsible
for determining the priority of each traffic category. If the backoff
counters of multiple traffic categories reach zero at the same time, there
is a {\em virtual collision} within the same physical node but different
traffic categories. This is resolved by a scheduler inside the node by
allowing transmissions from the traffic category with the higher priority.

Another enhancement in 802.11e is the concept of a {\em transmission
opportunity} (TxOP).  TxOP is defined as the interval during which a
node has the right to initiate transmissions. Thus, a node can initiate
multiple transmissions as long as its TxOP has not expired. The value
of EDCF-TxOP is unique throughout the network.

IEEE 802.11e also facilitates parameterized QoS. This is provided by the
enhanced version of PCF (EPCF). The HCF has a notion of Hybrid Coordinator
(HC) similar to the PC in PCF. The HC can allocate TxOP to itself or
any other node at any time but after sensing the channel to be idle
for a duration of PIFS, which is shorter than DIFS. This ensures that
the HC has the highest priority over all other nodes at any given time.
The HC allocates TxOP to pollable nodes in contention free periods and
sometimes even during contention periods. The main distinction between PCF
and HCF is that in PCF a node can transmit only one frame after receiving
CF-Poll frame. In HCF, a node can initiate the transmission of frames till
the end of TxOP. The duration of TxOP is notified to the node using the
Poll frame sent by the HC.  To provide TxOPs with appropriate duration
and appropriate time, the HC needs to obtain the pertinent information
from the individual nodes from time to time. For this purpose, the HC
initiates the so-called controlled contention periods during which nodes
send their resource requests to the HC without contending with other
nodes which are sending only data traffic.  There can be eight more
traffic categories for parameterized traffic in addition to the eight
EDCF categories. The frames queued in these categories are transmitted
after receiving poll frames from the HC. Thus, at most eight flows on
any node can be provided parameterized QoS. The flows are identified by
source and destination MAC addresses.

S. Mangold et al.~\cite{mangold} present a comprehensive overview of
these features.  They evaluate various QoS support features by means
of simulations. The simulations were mostly for 802.11a network (a
higher speed version of 802.11 operating at 5GHz band). They analyzed
the behavior of EDCF for different categories (mainly high, medium,
and low priorities)  and the measured throughput performance against
the offered load.  The simulation results obtained were consistent with
the expected results. For high priority traffic the throughput increased
linearly with the increase in offered load. Whereas, the low and medium
priority traffic observed knee points after certain level. The knee point
of low priority traffic appeared ahead of medium priority traffic. They
also evaluated the performance of HCF for parameterized QoS. The metric
chosen was the service delay for frames. The results were consistent with
expectations. The delay probability decreased for higher values. Almost
all the time the delay was within fixed bounds. The evaluations showed
that the enhanced MAC protocols indeed provide the expected QoS which
is the main objective of 802.11e.


\figw{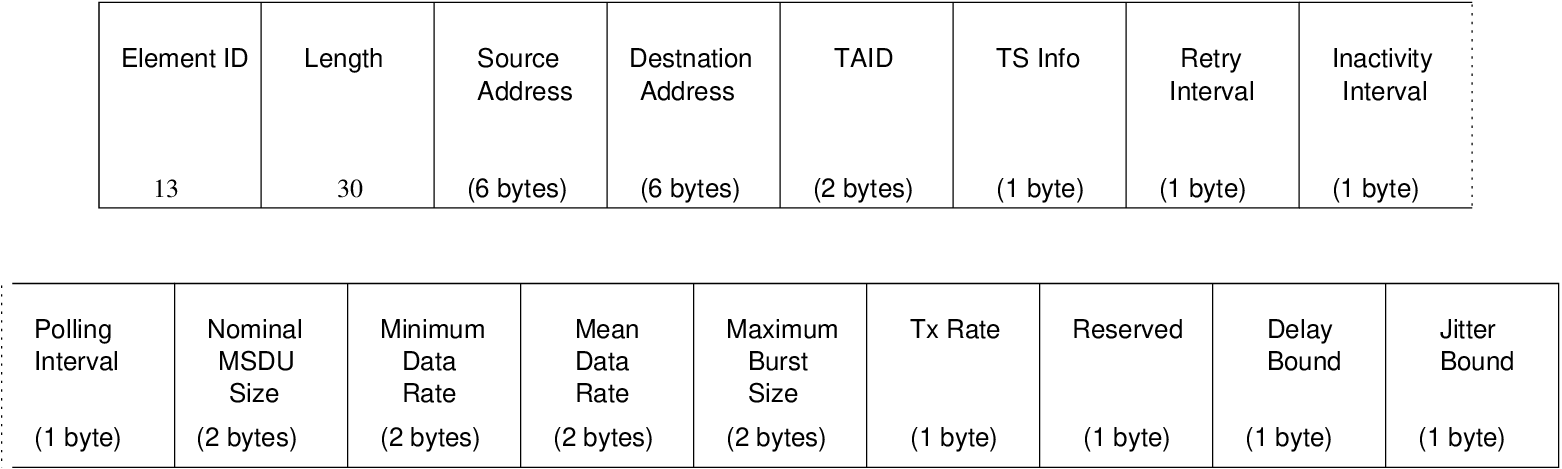}{6.2in}{\sl TSPEC element in 802.11e. This might be replaced
with a new Queue State element. The fields are mostly RSVP specific.}

One of the major issues with QoS provisioning is how to percolate
the QoS related information to pertinent layers. For example, the QoS
requirements are mainly application specific requirements, whereas,
the actual bandwidth provisioning has to be done at the MAC layer in
the LAN. This issue calls for a coordination between the MAC and higher
layers so that the applications can request their QoS requirement
in an appropriate manner. For this purpose, 802.11e defines two
entities; Station Management Entity (SME) and MAC Layer Management
Entity (MLME). SME is a logical entity in a node which is capable of
communicating with all layers in the network stack, whereas, MLME deals
with interacting with SME and managing MAC layer. The SME and MLME
communicate with each other by means of intra-STA signaling. Different
MLMEs in an infrastructure network communicate by means of inter-STA
signaling. For example, the communication between HC and a node regarding
TxOP duration etc. is part of the inter-STA signaling.  The inter-STA
signaling is used to setup, modify, and delete traffic streams. This
signaling carries a traffic specification (TSPEC) element to characterize
the QoS requirements of a node. As shown in Figure~\ref{tspec:fig}, TSPEC
carries various information like minimum data rate, burst size, etc.,
which can be derived directly from higher layer requirements. Whereas,
some fields like polling interval, retry interval, etc., are more MAC
layer specific.  Since SME is capable of interacting with all protocol
layers, it is capable of obtaining QoS requirement for specific flows.
These requirements are then conveyed to the MLME which is responsible
for establishment of final reservations. Sai Shankar et al.~\cite{sai}
describe the QoS signaling procedure for parameterized traffic in the
purview of RSVP.

As of beginning of year 2003, the 802.11e is still a draft and has not
been standardized yet. The QoS mechanism {\em seems} to be geared more
toward RSVP. (TSPEC is mostly RSVP specific). The draft is continuously
being revised. For example, there are some documents which indicate
that the TSPEC element is being replaced with a new Queue State
element~\cite{peter}.

\Se{Fairness}

Wireless channel is a shared scarce resource.  The MAC protocols used over
wireless networks are distributed protocols which try to avoid collisions
and provide the nodes in a network with an access to the channel in a
fair manner.  The efficiency of MAC protocols can be measured using two
parameters: the {\em probability of collision} and {\em fairness} in the
allocation of channel to competing nodes. The wireless LAN protocols,
like any other randomized multiple access protocols, try to resolve
the collision problem by following a {\em binary exponential backoff}
(BEB). BEB is a very efficient mechanism in terms of reducing collision
probability.  It often reduces the collision probability to a fraction
of transmissions, as low as 1\%.



Typically all variations of CSMA/CA protocol suffer from the {\em fairness
problem} investigated first by Bhargavan et al.~\cite{macaw}. A channel
access protocol (MAC) is termed to be unfair if it fails to provide the
channel access to individual nodes without giving preference to one
node over others when there is no explicit differentiation. That is,
when multiple nodes in a network are competing with each other for
channel access, the probability of each node winning the contention
should be equal.


Though the wired Ethernet protocol based on CSMA/CD is known to be
fair, its wireless counterpart 802.11b based on CSMA/CA is proven to
be unfair~\cite{koksal}. The unfairness of wireless networks has roots
in the fact that unlike wired networks, the collisions in wireless
networks are asymmetric. It is not necessary in wireless networks that
all nodes involved in collision suffer from packet loss.  The collisions
and hence the binary exponential backoff can occur primarily because of
the following three reasons.

\begin{itemize}

\item{Transmissions from two nodes interfere with each other and hence
their transmissions do not get acknowledged. The absence of ACKs is then
treated as collisions by both the senders.}

\item {In the 4-way handshaking mode, if a node does not receive CTS
response for its RTS request, it treats this as a collision and hence
doubles its backoff window.  This is irrespective of the status of the
destination node. A node may defer to send back CTS if any other node
in its vicinity has reserved the channel by sending an RTS or CTS to
some other node.}

\item{If two nodes carry out simultaneous transmissions intended
for the same destination, one of them may succeed because of
higher power level. This is called the capture effect in wireless
channels~\cite{zoran}.}

\end{itemize}

Out of these three collision scenarios only the first one is the real
collision. The other two scenarios result in success for one node and
failure for the other. In this case, only the failed one performs
the binary exponential backoff. For subsequent channel contention,
the node which succeeded recently has a higher probability of winning
the contention because of its lower backoff window. This is essentially
because of the dissimilar congestion view about the channel by different
nodes. Nodes which are generally successful in accessing the media
perceive it to be less congested compared to the nodes which encounter
failures. This prompts the successful nodes to access the channel in a
more aggressive manner (because of lower backoff window) than the failed
nodes. This skewed notion of congestion leads to an unfair access of
the channel.

The dissimilarity in wired and wireless networks arises because of the
dissimilar nature of the media. In wired networks the media is indeed a
shared media. If one node is accessing/using media all other nodes are
aware of the media access. But the wireless media is a {\em piecewise
shared} media. The reach of each node is limited by the transmission
power and the local noise present in the region. This makes the media
characteristics location dependent and hence a non-uniform nature of
the media is perceived by constituent nodes.

The backoff procedure used in almost all wireless medium access protocols
is essentially borrowed from the wired Ethernet where the non-uniform
nature of media does not exist. So, the binary exponential backoff
procedure which provides a fair media access in wired networks becomes
the cause of unfairness in the wireless networks.

\Sse{Impact of Unfairness}

The unfairness of MAC has a far reaching impact on the behavior of higher
layer protocols and the applications using the network. Application like
audio/video streaming are sensitive to packet delays and jitters. When the
underlying link behavior is unfair, some applications may be starved of
bandwidth just because their share is unfairly distributed somewhere else.

Shugong Xu et al.~\cite{xu} analyze the behavior of TCP protocol in
multihop 802.11 networks.  Using simulations they show that TCP suffers
from instability and unfairness problem in these networks. The instability
causes the throughput of available wireless network to fluctuate because
of interactions between different nodes carrying TCP-data and TCP-ACK
traffic. The unfairness problem leads to indefinitely long timeouts
causing multiple retransmissions and route breakups.

Koksal et al.~\cite{koksal} describe a scenario where TCP performance
degrades because of short term unfairness exhibited by MAC
protocols. Because of short term unfairness the TCP acknowledgments
fail to reach the sender in a timely fashion. This results into a
bursty traffic. This bursty traffic results into and ACK compression
which aggravates the burstiness of the stream. The bursty traffic has
many disadvantages like packet loss in response to bursty traffic and
throughput loss because of idle links during two consecutive bursts.

\Sse{Achieving Fairness}

Since the fairness problem is deep rooted in the MAC layer itself, it is
reasonable to conclude that it can be solved by modifying the MAC in an
appropriate way that achieves fairness. To this effect there have been
several enhancements and modifications suggested to the MAC layer. Some of
the examples are MACAW~\cite{macaw}, Estimation based backoff~\cite{fang},
Distributed Wireless Ordering Protocol~\cite{kanodia}, and Distributed
Fair Scheduling~\cite{vaidya}.

\begin{figure*}
\begin{minipage}[t]{3.2in}
\centerline{\psfig{figure=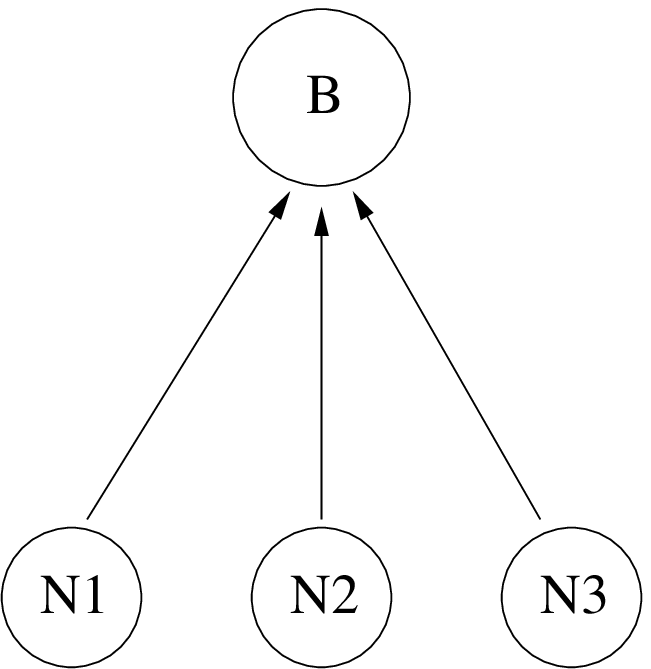,height=1.25in}}
\caption{\small{\sl A single cell configuration. All nodes are within range of
each other. Nodes N1, N2, and N3 contend for channel to send data to node B.
If some node has very small backoff window than others, it may end up grabbing the channel
and as a result other nodes may fail to get their fair share.}}
\label{cell:fig}
\end{minipage}
\hspace{.05in}
\begin{minipage}[t]{3.2in}
\centerline{\psfig{figure=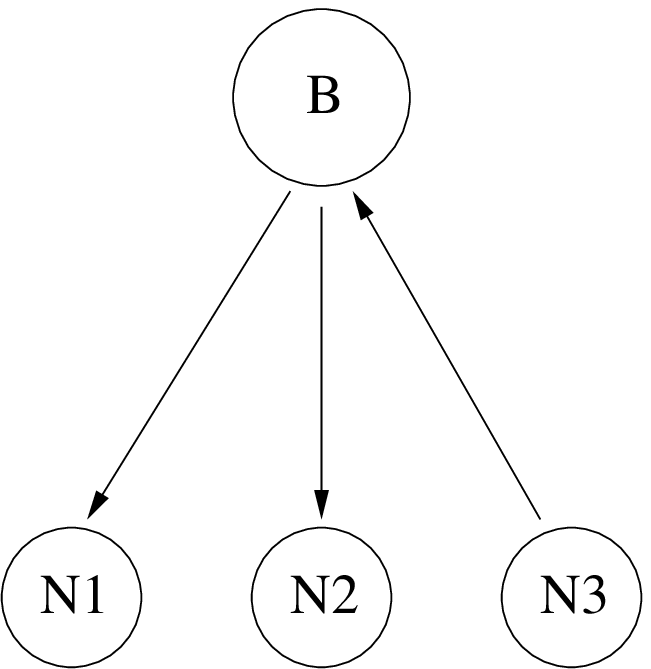,height=1.25in}}
\caption{\small{\sl Node B is sending data to node N1 and N2. Node N3 is sending data to
B. If the fairness criteria is dependent on share per node rather than share per stream,
then nodes like base stations are starved for bandwidth as these are the nodes which 
have maximum number of streams for downstream traffic.}}
\label{multistream:fig}
\end{minipage}
\end{figure*}

\paragraph {\bf MACAW\\} Vaduvur Bhargavan et al.~\cite{macaw} observe
that to allocate media fairly, congestion level estimation should be
a collective effort. In other words, the propagation of congestion
information should be explicit rather than each node learning it on
its own.

MACAW tries to address the fairness issue in single cell scenarios as
depicted in Figure~\ref{cell:fig}.  In this setup there are 3 nodes N1,
N2, and N3 which are trying to transmit data to another node B (base
station) in same cell. If the transmission queues of all the nodes are
backlogged then most of the times all of them are contending for the
channel access. If during the contention all but one pad have relatively
high backoff counters then the one with low backoff counter will win
the contention. This will result in that node resetting its contention
window to the lowest possible window size. This will put the successful
node at an advantage over the other nodes since it has now a more than
fair chance of winning the contention at later time.  This will result
in increasing the contention windows of other even further. Note that,
in this scenario the implicit assumption is that the other nodes {\em
do not perform carrier sense and do not freeze} their backoff counters
for the duration of transmission.

The observation to be made here is that the backoff counter (or the
contention window) of a node does not reflect the ambient congestion
level that exists in the entire network. MACAW advocates sharing of
this information by all the nodes in the network.  This can be done by
including the value of current contention window in every packet that
gets transmitted. All other nodes can now update their contention values
depending on the latest notified value. Thus, in a network where all
nodes are within the reach of each other, after every transmission all
nodes have same contention window.

This scheme solves the problem of information sharing but introduces a
new problem of oscillating contention windows. After every successful
transmission, the contention window gets reset to the minimum possible
value. If the number of nodes in a network is very high, all nodes would
spend a lot of time adjusting their contention window to reach an optimal
level which would be immediately reset after a successful transmission
by any of the nodes. MACAW deals with this oscillation problem by
introducing a {\em multiplicative increase linear decrease} (MILD)
backoff algorithm. In this algorithm the backoff window is increased
by a multiplicative factor after collision and is decreased by 1 after
success. The multiplicative increase yields a prompt convergence to the
backoff window when the contention is high and avoids oscillations by
not resetting it to the minimum possible value.

MACAW does not assume that the nodes use carrier sense to avoid
collisions.  In a single cell network, the fairness problem {\em does not
exist} if nodes perform carrier sense and freeze their backoff counters
for the duration of transmission by other nodes.

MACAW also tries to address the issue of proper definition of
fairness. If the fairness criteria is defined as,``equal channel share
for every node,'' then there are certain nodes like base stations in
infrastructure networks which are at a disadvantage.  Consider the
setup shown in Figure~\ref{multistream:fig}. Here the base station B
needs handle streams to nodes N1 and N2. Node N3 needs to send data
to the base station. In this scenarios there are only two nodes which
need to access the channel. If the channel access is granted equally
to N3 and base station then the channel share is not fair from the
perspective of streams. MACAW suggests that this problem can be solved
by running multiple instances of backoff algorithm each corresponding
to a stream. As a matter of fact, this scheme is the basis of traffic
categories in 802.11e.

\paragraph {\bf Estimation based backoff\\} Zuyuan Fang et al.~\cite{fang}
propose a novel measurement and estimation based scheme to solve the
fairness problem in 802.11b networks. The proposal is based on estimating
the throughput of all nodes and then computing a {\em fairness index}
based on this estimation. This fairness index can then be used to adjust
the contention window.

The fairness index between any two nodes $i$ and $j$ is computed using the
following equation:

\begin{equation}
fairness~index = max  \{{\forall}{i,j,}~{min_{i,j}(\frac{W_i}{{\phi}_{i}},\frac{W_j}{{\phi}_{j}})/max_{i,j}(\frac{W_i}{{\phi}_{i}}, \frac{W_j}{{\phi}_{j}})} \}
\end{equation}

where $\phi_i$ is the predefined fair share that a station {\em i} should  receive and
$W_i$ is the throughput achieved by node {\em i}. The constraints on $\phi_i$ and 
$W_i$ are such that for a total throughput of W :\\
{$\forall$$i$ $\sum$ $\phi_i$ = 1} and {$\forall$$i$ $\sum$ $W_i$ = W}

The fairness goal now becomes maximizing the value of {\em fairness
index} locally.  When the fairness index equals 1 all nodes obtain the
throughput proportional to their share. There are two issues with this
approach. First, how to predefine the fair share of each node. Second,
how to measure the total usage by all nodes. Note that the algorithm
should take into account the hidden nodes and node mobility as well. Thus,
there cannot be any signaling traffic between the nodes.

The solution to the first problem is achieved by dividing the nodes
into two partitions. In this approach, each node regards all of its
neighbors as a single entity with a notion of {\em myself} and the {\em
others}. Now the fair share assumed by each node is $\phi_i$ = 0.5. All
other nodes also get the remaining share of 0.5. The assumption here is
that when all nodes use the same approach, all of them will compete with
same fair share. Thus due to the local contention and collisions the long
term throughput will achieve fairness. If one wants to provide a stream
based fairness, a node can increase its fair share by the proportional
amount. That is, if a node has two streams then it treats streams of all
other nodes as a single stream and computes its fair share as $\phi_i$
= 0.67.

The second problem is solved by snooping on the traffic in the network. A
node can always measure its own traffic. The traffic from hidden nodes
is deduced by snooping on the CTS and ACK packets sent to those nodes.

Once a node has an estimate of its own traffic $W_{ei}$ and the traffic by
others $W_{eo}$ it computes the fairness index using the above mentioned
equation. After computation of the fairness index it doubles it contention
window if it has obtained more than its fair share and halves it if it
has not received its fair share.

The effectiveness of this approach essentially depends on the
approximation of the fair share $\phi_i$. if the value of $\phi_i$
is higher than what it ought to be, the node acts in a greedy fashion
and the short term fairness of the channel allocation is affected. For a
network with $n$ competing nodes, the share should be $1/n$. The farther
the assumed value (0.5) from $1/n$, the greedier is the algorithm. This
greediness of algorithm, though ensures the long term fairness, affects
the short term fairness severely. Moreover, the IEEE 802.11b MAC is known
to be fair in long term and it is the short term fairness one needs to
address. This can be possibly done by estimating the number of nodes in
the network. If the estimation of number of nodes is close to the actual
number of nodes, the algorithm can work very efficiently.


\paragraph {\bf Distributed Fair Scheduling\\} Nitin Vaidya et
al.~\cite{vaidya} extend the notion of fairness to include the weight
of a flow for considering its share of the channel. This is in contrast
with the prevalent notion of equal share to all nodes or equal share
to all flows. They propose an extension to 802.11b DCF to achieve this
Distributed Fair Scheduling (DFS) with weighted proportions.  The DFS
was designed in an attempt to emulate the Self-Clocked Fair Queueing
(SCFQ)~\cite{scfq}. The DFS is specifically tailored for distributed
systems like 802.11b LANs.

\begin{figure*}
\begin{minipage}[t]{3.2in}
\centerline{\psfig{figure=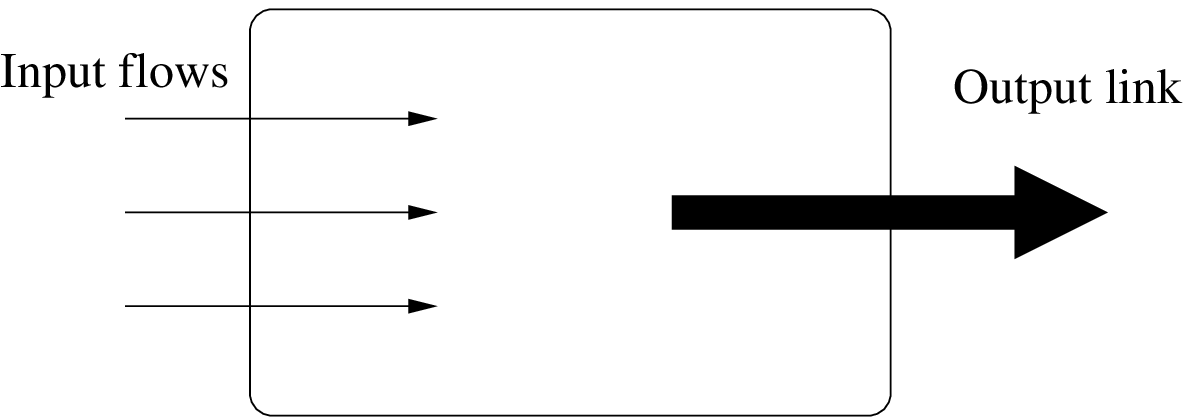,width=3in}}
\caption{\small{\sl The SCFQ algorithm is used for scheduling packets from multiple
input flows to one output link. The DFS uses this as the basic model.}}
\label{scfqlink:fig}
\end{minipage}
\hspace{.05in}
\begin{minipage}[t]{3.2in}
\centerline{\psfig{figure=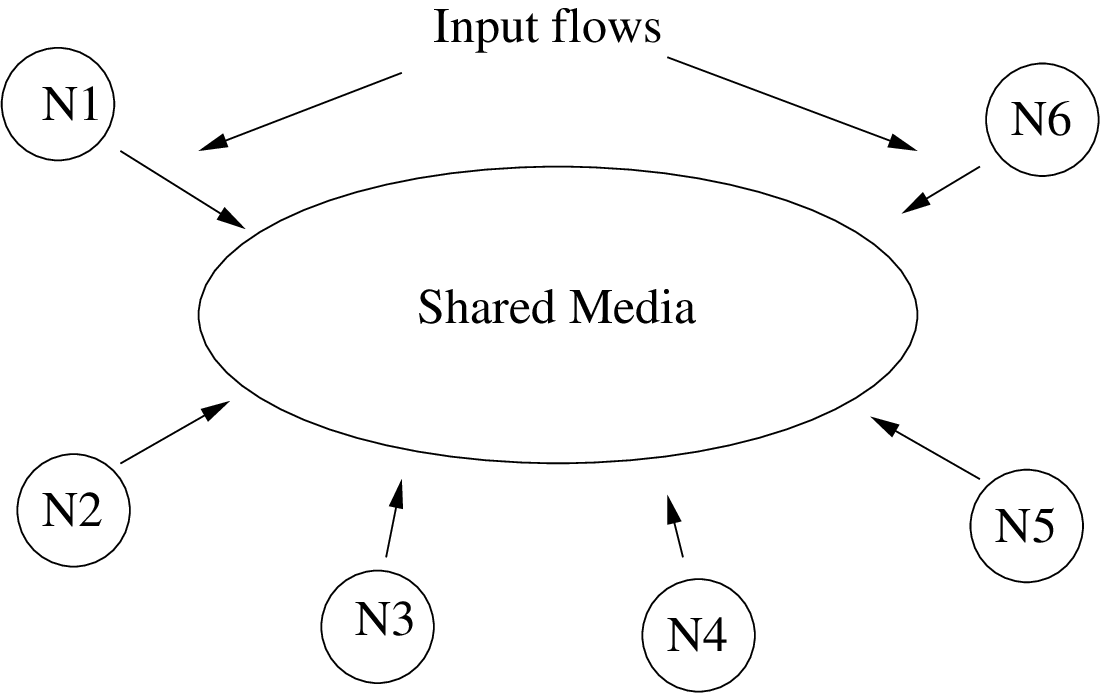,width=2.5in}}
\caption{\small{\sl  The DFS maps the link in SCFQ to the shared medium and the  input
flows to the traffic from wireless nodes. DFS is thus a distributed version of SCFQ}}
\label{dfsshared:fig}
\end{minipage}
\end{figure*}

SCFQ is a centralized algorithm for packet scheduling on a link shared
by multiple flows as shown in Figure~\ref{scfqlink:fig}. The central
coordinator maintains a virtual clock.  At any given time {\em t},
{\em v(t)} indicates the {\em virtual time}.  Assume that : $P^k_i$
denotes the $k^{th}$ packet arriving on $i^{th}$ flow, $A^k_i$ is the
real time of arrival for packet $P^k_i$, and $L^k_i$ represents the size
of the packet $P^k_i$. For each packet $P^k_i$, a start tag $S^k_i$ and a
finish tag $F^k_i$ are assigned. The assignment algorithm is as follows:\\

\begin{enumerate}

\item $\forall$ i $F^0_i$ = $0$.

\item Every packet $P^k_i$ is stamped with start tag $S^k_i$ using following
equation:\\
	$S^k_i$ = max \{v($A^k_i$), $F^{k-1}_i$\}.
	
\item A finish tag for each packet $P^k_i$ is calculated as:\\
	$F^k_i$ = $S^k_i$ + $\frac{L^k_i}{\phi_i}$ .

	where $\phi_i$ is the bandwidth share of the flow {\em i}.

\item Initially at time $t$ = 0, the virtual clock {\em v(0)}  is set to 0.
	The virtual time is updated after every packet is transmitted. At the end of
	the transmission of packet $P^k_i$, the virtual clock is set to $F^k_i$.

\item Packets are selected for transmission in an ascending order of finish tags.

\end{enumerate}

Alternatively one can assign the start tag for a packet based on
the real time when it is advanced to the front of the queue. If the
packet arrives when the flow is empty the the time $f^k_i$ = $A^k_i$
is the real time when the packet is advanced to the front of the queue,
else $f^k_i$ is the real time when the packet $P^{k-1}_i$ finishes the
transmission. Thus the start tag for a packet can be assigned in a lazy
fashion by following equation\\

\begin{equation}
S^k_i = v(f^k_i)
\end{equation}

Like SCFQ, DFS determines the packet transmission times based on the
finish tag of each packet. Further, the virtual time is updated in
the same way as SCFQ.  DFS tries to map the shared wireless medium
to the output link and input flows paradigm of SCFQ. As shown in
Figure~\ref{dfsshared:fig}, the output link in Figure~\ref{scfqlink:fig}
is mapped to the shared wireless medium and the input flows are mapped
to the flows from each individual node.

SCFQ is a centralized algorithm. Because of this centralized nature, there is no issue
in determining the shortest of the finish tags of the packets from each queue. But in
DFS all queues are distributed. Thus, the selection of next eligible packet for
transmission has to be done in a distributed manner. DFS circumvents this problem by
choosing the backoff interval proportional to the finish tag of the packet that is at
the front of the flow. Further, each transmitted packet carries the virtual finish time
of the packet each transmitted packet carries its virtual finish time with it. This
virtual finish time is used by other nodes to synchronize their virtual clocks.

DFS is implemented as follows:\\
\begin{itemize}
\item Whenever a packet advances to the front of the flow, its start tag is updated
depending on the virtual time at that instance.

\item The virtual finish time for every packet is calculated similar to SCFQ but with
a scaling factor for choosing a suitable scale for virtual time.
Note that $S^k_i$ = v($f^k_i$) and from equation (2):\\
\begin{equation}
F^k_i = v(f^k_i) + scaling~factor *\frac{L^k_i}{\phi_i}  
\end{equation}

\item Using this virtual finish time a backoff interval is picked for the packet $P^k_i$:\\
$B_i$ = $\lfloor$ $F^k_i$ - v($f^k_i$) $\rfloor$\\
Using equation (3) we get:
\begin{equation}
B_i = \lfloor scaling~factor * \frac{L^k_i}{\phi_i} \rfloor
\
\end{equation}

\item This backoff window $B_i$ is further randomized by multiplying it with a random
variable with mean 1.
\end{itemize}

On close observation, it becomes apparent that there is no need to maintain a virtual
clock as the virtual finish time is never used in calculating the backoff interval.
After every packet transmission on LAN, all nodes calculate the backoff interval based
on the fairness share of each node and the packet size. DFS deals with collisions the
same way as 802.11b, i.e., the binary exponential backoff. Thus only first backoff
interval is chosen depending on the fairness share of the node and is a {\em linear
function} of packet size and fairness share.  If the fairness share $\phi_i$ is very
small, there might be long duration of idle time because of this linear nature of
backoff interval. If the number of nodes in the network is high, this might lead to
dropped throughput of the network because of excessively long backoff intervals. This
problem is circumvented by compressing larger backoff intervals into smaller {\em
exponential range}.  It is not clear how the channel shares $\phi_i$ is assigned to
each node.


\Se{Performance}


Throughput performance of communication links is measured in terms of observed data
rates. Looking from the link layer perspective, the performance is the effective data
rate available for the raw bits exchanged between nodes. Whereas, looking from the
network layer perspective, it is the rate at which the network layer data is exchanged.
This precludes the management data exchanged at the link layer. As one moves higher up
the protocol stack, the performance of a link progressively reduces. The reason for
this progressive performance reduction is almost always associated with lower level
protocol overheads. Further, for shared media like wireless channels, where multiple
nodes contend for the same channel, additional channel bandwidth is used up to resolve
the contention and some bandwidth may even be lost when collisions arise.

\Sse{Performance Loss} 

For wireless networks like 802.11b, it is {\em never} possible to achieve the
theoretical available bandwidth at the application level. Since a wireless channel can
never be devoid of noise, the first loss is at the physical radio link itself.
Shannon's equation gives the relation between available data rate $R$ for a channel
with bandwidth $B$ with a Signal to Noise Ratio (SNR) as :\\

{\centerline {${R}$~=~${B}~{log_2}{(1+SNR)}$} } 

This is the theoretical limit and there can be additional loss because of {\em
multipath fading}, {\em Doppler shifts}, and {\em bit errors}~\cite{hpcn}.

The second rung of reduction in throughput occurs at the Physical layer. An 802.11b
frame is shown in Figure~\ref{80211frame:fig}. Each frame comprises of 24 bytes of
Physical Convergence Protocol Layer (PLCP) preamble and header. This PLCP preamble is
then augmented with rest of the MAC Protocol Data Unit (MPDU). The header is always
transmitted at a slower rate of 1 Mbps and rest of the MAC Protocol Data Unit is
transmitted at a variable rate. This slow rate transmission is necessary for
compatibility and reachability with all nodes in the network. Thus, a significant
portion of transmission occurs at the lower data rates causing a reduced throughput.

\figw{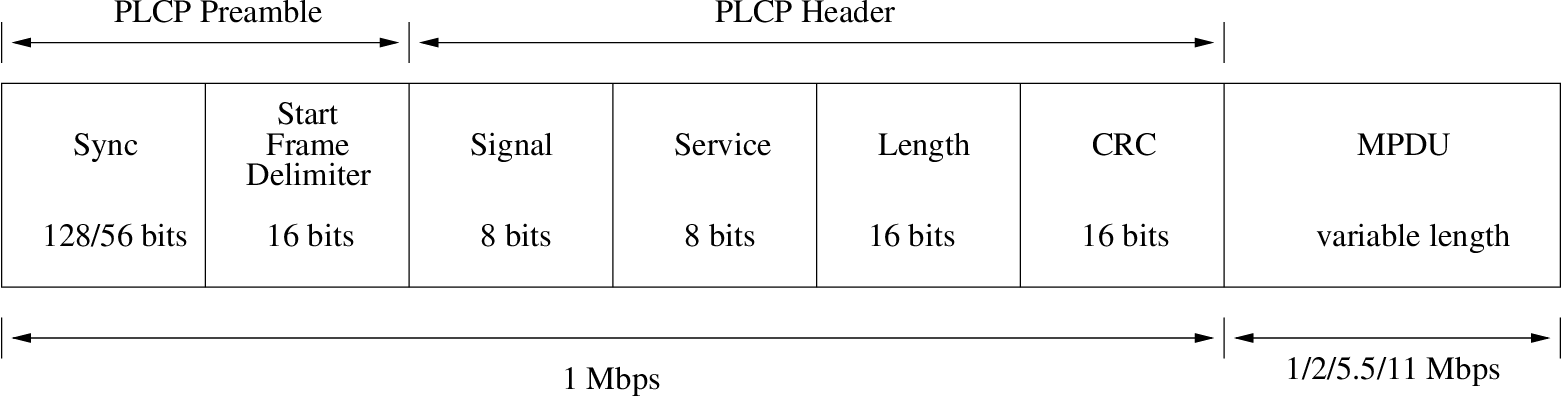}{6.2in}{\sl IEEE 802.11b frame format. Initial 24 byte header is always
transmitted at 1 Mbps. Rest of the MAC Protocol Data Unit (MPDU) is transmitted at
either 2 Mbps or 5.5 Mbps or 11 Mbps.}

IEEE 802.11b uses CSMA/CA~\cite{maca} protocol for media access. In this protocol each
data packet transmission is preceded by a channel reservation request (RTS) and a
channel reservation response (CTS) between sender and receiver. The data packet is then
followed by an acknowledgment from the receiver (ACK). The ACK is required for the
reliability of link level transmissions. The transmission follows the so called {\em
4-way handshake protocol} of RTS--CTS--DATA--ACK in order to solve the {\em hidden node
problem}.  The additional bandwidth consumed by these RTS/CTS/ACK and several other
management frames add up to the reductions in performance throughput.

The observed data rate for 802.11b networks by network layer is around 7 Mbps in
typical indoor environments and the bandwidth of 802.11b networks is 11 Mbps. The next
level of performance loss is because of complex interactions between protocol layers
and other overheads. We will be examining some of the protocol performance issues.

\Sse{Performance of TCP}

Performance of TCP/IP over wireless networks is a widely researched topic.  TCP
behavior is studied under various flavors of wireless environments.  The primary thrust
of this research is on the behavior of TCP in response to the error conditions in
wireless networks. TCP, which is known to be a very stable and robust protocol on wired
networks does not perform as well on wireless networks. Some of the serious issues with
TCP are that of performance degradation over wireless links. The primary reason for
performance degradation is the assumption by TCP that all losses in the network are due
to congestion~\cite{xylomenos-ieee}.

In almost all wireless networks, the transmissions are highly prone to frame errors
which increase with packet size. For example, in 802.11 networks, the Frame Error Rate
(FER) doubles for every 300 byte increment in frame size~\cite{nguyen}. Thus, at higher
packet sizes, the frame errors are higher and hence the packet loss is higher. This
packet loss is wrongly treated as congestion by TCP and its congestion control
mechanism is triggered.  As a result, TCP reduces the congestion window size. The TCP
congestion window size is the number of packet that can be sent with outstanding
acknowledgments. Congestion window is the measure of minimum number of packets required
to keep the link occupied increasing the channel utilization. This congestion control
causes lower channel utilization even though in reality there is no congestion in link.

One of the first proposed solutions for TCP performance degradation on wireless links
is {\em Indirect-TCP} (I-TCP) by A. Bakre et al.~\cite{itcp}. I-TCP works by splitting
the transport connection at the wired-wireless boundary, usually on base stations.
I-TCP maintains two separate TCP segments, one is over the wired network between the
base station and the wired end of TCP connection, the other is over the wireless
between the base station and the wireless host.  This way, the losses occurring on
wireless network are hidden from the wired nodes.  To cover up for the losses on
wireless segment the base station carries out retransmissions. Though an efficient
solution, this approach violates the end-to-end semantics of TCP.  For example, in this
approach even if the sender receives an acknowledgment, it does not mean that the
receiver has indeed received the packet. This violation of end-to-end semantics can
have serious impact on applications. It may so happen that the wireless host may crash
before it receives the packet that has been acknowledged by the base station. This is
against TCP semantics and may not be acceptable to certain applications.

Hari Balakrishnan et al.~\cite{bala} try to address this problem by proposing a {\em
snoop module} that resides in the routing protocol stack of the base station. This
snoop module snoops on all the packets that are sent to the mobile hosts and the
acknowledgments that are transmitted by the mobile nodes. It also caches the packets
that are transmitted over the wireless network. This enables the base station to detect
packet losses in the wireless network and to retransmit the lost packets, thus avoiding
the triggering of TCP's congestion control and avoidance mechanisms.

The 802.11b provides a fragmentation oriented solution for this at the MAC layer.  
Since the frame error rate increases with frame size, the reliability of transmissions
can be increased by reducing the frame size. Al most all 802.11b network interface
cards support an option of enabling the fragmentation. In the event of bad channel
conditions and increased frame errors, the MAC unilaterally fragments the larger frames
into smaller frames of optimal size. This is done in a manner transparent to the higher
layers. The receiving node then defragments the frames to obtain a complete data packet
which can be handed over to the higher protocol layer. Since the fragmentation occurs
at the frame level, there is no higher protocol overhead associated with the fragments.
There is a small overhead of additional MAC header. Compared to the alternative of
loosing the frame and reducing the overall throughput, a small overhead is a fair trade
off.

Another problem with TCP in wireless LANs is that of {\em Self Collision}. George
Xylomenos et al.~\cite{xylomenos} describe a scenario where the TCP data packets
contend for the channel with the ACKs for the previous packets. This contention results
in self collision for the TCP connection causing overall performance to degrade. This
self collision is caused because of the half duplex nature of wireless networks. Haitao
Wu et al.~\cite{haitao} propose a modification (described in section~\ref{Performance
Improvements}) to the Distributed Coordination Function (DCF), which is the MAC
protocol for 802.11b, to alleviate this self collision problem.


Typically most of the solutions for the issues faced by higher protocols like TCP are
localized solutions. The problems are tackled within a protocol layer and no effort is
made to make other protocol layers aware of the changing situation. While this
transparent approach is good from modularity and simplicity of design point of view, it
severely impacts the performance. It is evident that in order to optimize the network
performance, all layers in the protocol stack should adapt to the variations in the
wireless link appropriately. Further, this should be done while considering the
adaptive strategies at other layers~\cite{hpcn}.

\begin{table*}
\centerline{
\begin{tabular}{|c|c|c|c|c|} \hline
{\bf Data Rate} & {\bf Code Length} & {\bf Modulation} & {\bf Symbol Rate} & {\bf Bits/Symbol} \\ 
\hline
{1 Mbps} & {11 (Barker Sequence)} & {BPSK} & {1 MSps} & {1} \\ \hline
{2 Mbps} & {11 (Barker Sequence)} & {QPSK} & {1 MSps} & {2} \\ \hline
{5.5 Mbps} & {8 CCK} & {QPSK} & {1.375 MSps} & {4} \\ \hline
{11 Mbps} & {8 CCK} & {QPSK} & {1.375 MSps} & {8} \\ \hline
\end{tabular}
}
\caption{\small {\sl Different possible data rates for 802.11b networks.}}
\label{rates:tab}
\end{table*}

\Sse{Performance Improvements}

The performance improvement research for IEEE 802.11b can be broadly categorized into
three classes based on the locality of the optimizations involved.  These categories
are mainly: MAC performance optimization, Network or Transport layer enhancements, and
Infrastructural arrangements.

\Ssse{MAC Performance Optimization} 

Recent performance optimization research in 802.11b MAC aims at improving or enhancing
the performance by modifying the MAC while retaining backward compatibility with
existing specification. We give an overview of some schemes seeking to modify the DCF
channel access protocol to bring in performance enhancements.

\paragraph {\bf Multi-Rate 802.11b\\} IEEE 802.11b supports data transmission facility
at multiple rates. These multiple data rates are possible because of different
modulation techniques which are optimized for different channel conditions.  
Techniques like Quadrature Phase Shift Keying (QPSK), with 8 bit Complementary Code
Keying (CCK) error correction codes, provide a bandwidth of 11 Mbps but need very high
Signal to Noise Ratio (SNR) in channel.  Whereas, techniques like QPSK, with 11 bit
Barker Sequence, provide 2 Mbps of bandwidth but are capable of operating in noisy
environments. Table~\ref{rates:tab} describes various data rate specifications for IEEE
802.11b.

Network interface Cards supporting these multiple data rates are capable of switching
between different modulation techniques after assessing the channel characteristics.  
A user can choose to clamp the network interface at one particular rate or can enable
the option of letting the device choose an appropriate rate. In order to improve the
performance, the network interface cards can adapt to varying channel conditions and
dynamically switch between different modulation techniques. This adaptation primarily
involves two tasks, (1) sensing the channel quality and (2) selecting the appropriate
technique and hence the data rate.  Channel quality can be estimated by using several
metrics, such as, signal to noise ratio, bit error rate, signal power, etc. A history
of these metrics can be maintained which can be used to predict the channel conditions
in future. But given the volatile nature of wireless channels, it is not clear how
accurate and reliable these estimates can be. Using these estimates, one can select an
appropriate rate for the estimated channel condition and transmit the data at the
selected rate. For example, if the channel condition is excellent then the sender can
select the highest possible rate for transmissions. On the other hand, if the bit error
rate is high, the sender can choose to select a stronger encoding technique while
dropping the rate. For worse channel conditions, the sender may resort to the lowest
possible rate just to get the data across.

\paragraph {\bf Auto Rate Fallback Scheme (ARF)\\} The channel quality estimation is a
proactive approach and may not be accurate. Instead, one can implement a simple
reactive channel quality sensing mechanism which gauges the changing channel conditions
based on success or failure of previous transmissions.

In IEEE 802.11b, link level reliability is provided by explicit link level ACKs.  
Retransmissions are carried out if positive acknowledgments are not received.  Lucent
corporation's Orinoco cards provide a simple result based Auto Rate Fallback
(ARF)~\cite{arf} mechanism which uses ACKs (or their absence) to estimate the channel
quality. The ARF mechanism is a timer driven mechanism which keeps track of missed
acknowledgments and works as follows. When an ACK is missed for the first time, after
earlier successful transmissions, the first retransmission is carried out at the same
rate. After second failure, the transmission rate is downgraded to the next lower data
rate and a recovery timer is started.  The transmission rate is upgraded back to
the next higher data rate if the timer expires or 10 consecutive transmissions are
successful. After this recovery if the very next transmission meets a failure, the
system immediately reenters the fallback condition and resumes the normal operation.

While the ARF mechanism is good for link quality estimation between a fixed pair of
nodes, it overlooks the fact that it is the {\em receiver} whose channel conditions
need to be estimated and not the sender. A direct disadvantage of ARF scheme can be
seen when there are multiple nodes communicating with each other in a wireless network.  
If a node moves to a location with bad radio characteristics, other nodes communicating
with this particular node would experience transmission failures and as a result the
transmission rate would be dropped. Consequently, it would take 10 successful
transmissions or a timeout of the recovery timer to increase the data rate to the next
higher level. This can result in pulling down the throughput for the entire network at
lower data rates. If more than 10\% of the nodes in a wireless network are in bad radio
characteristics zone and the communication traffic between all nodes is more or less
equal, the overall throughput of the network would always be clamped at the lower
threshold. In addition, if some nodes are mobile and there are certain dark spots in
the wireless network, the nodes moving in and out of dark spots would reduce the
network throughput further. Thus, {\em ARF would fail to perform efficiently in a
network which comprises of zones with varying network characteristics}. Which defeats
the primary goal of ARF, that is to adapt to varying network conditions.

\paragraph {\bf Receiver Based Auto Rate Scheme (RBAR)\\} Gavin Holland et
al.~\cite{rbar} propose a rate adaptive scheme which uses feedback information from the
{\em receiver} to sense the channel condition rather than estimating it at the sender.
They also observe that in cellular networks, most of the receiver based channel quality
estimation techniques have following characteristics:

\begin{itemize} 

\item {Channel quality information is estimated by the receiver and periodically fed
back to the sender either on same channel or on a separate channel.}

\item{The sender performs rate selection based on the feedback from the receiver.}

\item{The estimation, feedback, and rate selection schemes often reside at the physical
layer and are transparent to the higher layers (such as, even MAC).}

\end{itemize}

It is difficult to apply same techniques to wireless LANs because conventional wireless
LANs usually operate in half duplex mode\footnote{For full duplex mode, one would need
the receiver and transmitter on same card to operate simultaneously. The transmitter in
this case would interfere with the ongoing packet reception. Hence most of the wireless
LANs operate in half duplex mode.}. This makes simultaneous feedback impossible.
Moreover, wireless LANs use distributed contention based media access which need
accurate time estimates for packet transmissions so that other nodes in the network can
be informed about them. Dynamic encoding below MAC layer would hinder this operation
and can impact efficiency.

Thus, one of the major problems in receiver based estimation schemes for wireless LANs
like 802.11b is the requirement of a mechanism to provide feedback information to the
sender. Another desired feature about the feedback is that it should convey
instantaneous information rather than history to be more effective.

Since, IEEE 802.11b uses DCF for channel coordination, each DATA packet from sender is
preceded by a short RTS packet.  RBAR leverages on this RTS packet to estimate the
channel quality and reception at the receiver end. Using this information the receiver
selects an appropriate transmission rate. This rate information is then piggybacked on
the CTS response to the sender. The sender then uses this information to carry out the
actual transmission.

\figw{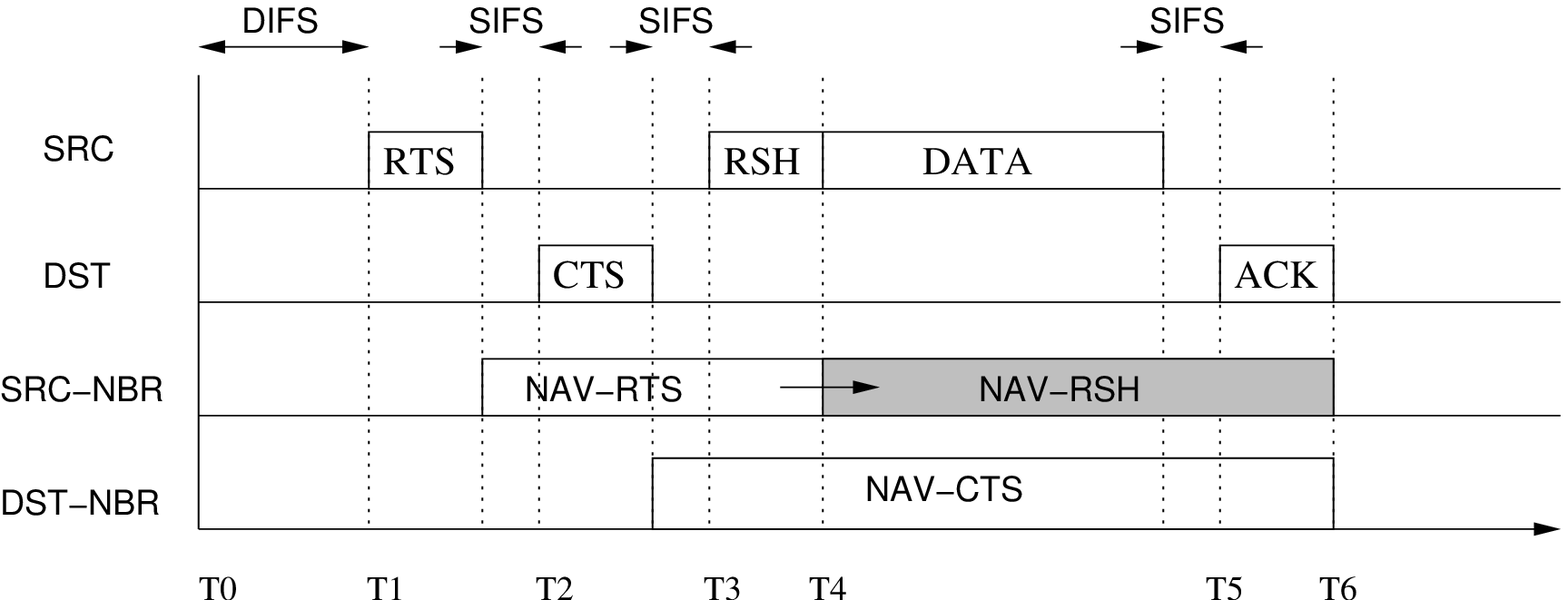}{5.5in}{\sl Message exchanges for RBAR. Each DATA packet is preceded by RTS
and CTS messages. On hearing RTS, the nodes in the vicinity of sender set their NAV to
the duration mentioned calculated using the tentative rate and size. The receiver
analyzes RTS packet and selects an appropriate rate and responds with CTS message. On
hearing CTS, the nodes in the vicinity of receiver set their NAV to the duration
calculated in the same way. This causes in establishment of channel reservation till
the time the ACK is sent back to the sender. If the selected rate is different than
tentative rate, the sender prepends a Reservation Sub Header (RSH) to enable nodes in
its vicinity to update their NAVs}

In 802.11b DCF, RTS and CTS packets carry the {\em duration} information for which all
nodes should set their NAV. RBAR proposes to replace this duration information with
data rate and the size of the data packet. Other nodes overhearing RTS and CTS can
calculate the duration from this information and still set their NAVs to appropriate
values.

The operation sequence and the 4-way handshaking for RBAR is shown in
Figure~\ref{rbar:fig}. Whenever some node {\tt Src} has some data to be sent to some
node {\tt Dst}, it performs the usual channel sensing for a period equal to DIFS. Once
the channel is sensed to be idle for DIFS time it sends a short RTS message to {\tt
DST}. This RTS message carries the tentative transmission rate and the size
information. Using this information, nodes in the vicinity of {\tt Src} calculate the
duration and set their NAVs accordingly. The node {\tt Dst} upon receiving RTS analyzes
it for its signal quality, strength and other metrics and selects an appropriate rate
for the transmission. This selected rate and the size of data is again encapsulated in
the CTS response. Nodes overhearing CTS response set their NAVs to appropriate values.
The node {\tt Src}, after receiving CTS response, selects the suggested rate and
carries out the data transmission. In the event that the selected rate is different
than the tentative rate, the nodes in the vicinity of {\tt Src} who can not hear the
CTS response also need to be informed of the change in the duration. For this purpose,
RBAR prepends the DATA packet with a {\em Reservation Sub Header} (RSH) which informs
other nodes of modified rate and size. This is used to modify the NAVs of hosts in the
vicinity of {\tt Src}.  This way the data transmission is carried out with the rate
suggested by the sender.

The overhead of RBAR comes in the form of the additional RSH frame. This frame is
required when the actual transmission rate is different than the tentative transmission
rate. The overhead is maximum when the data packet size is the smallest and decreases
with an increase in the data packet size. The authors analytically show that even for
small packets with size of 32 bytes RBAR outperforms ARF scheme by at least 10\% in
terms of the overall channel throughput. With increasing packet sizes, gains up to 20\%
are observed.

The gains are significant and the required modifications are relatively trivial. This
makes RBAR approach a very elegant solution to deal with varying channel conditions.
But RBAR heavily relies on a single RTS frame to infer the channel conditions and
select the optimal rate. Further, in 802.11b, the RTS frame is always transmitted at
the lowest possible rate so that all nodes can update their NAVs. Thus, a single RTS
frame may be inadequate to infer proper channel characteristics.

\paragraph {\bf Opportunistic Auto Rate Scheme (OAR)\\} B. Sandeghi et
al.~\cite{sandeghi} propose an opportunistic media access scheme (OAR) which extends
RBAR and tries to utilize the period of good channel conditions to maximize the network
throughput. It exploits the fact that the {\em coherence period} for 802.11b networks
is in the range of multiple milliseconds even at mobility speeds of 20 m/s.  The {\em
Coherence period} of a channel is the interval for which the SNR values do not
decorrelate and the channel quality remains fairly unchanged. Thus the channel quality
estimation done at an instance can be assumed to be valid for the duration of the
coherence period. For 802.11b, this period corresponds to multiple packet transmission
times.  This enables OAR to rely on the signal quality information provided by the
receiver and carry out multiple packet transmissions.  The main idea of OAR is to
exploit good channel conditions to transmit as many packets as possible while retaining
the long term fairness provided by 802.11b. OAR achieves this by sending a burst of
packets for a single RTS-CTS handshake. For this purpose, OAR suggests a modification
to the 4-way handshake mechanism, which can be retrofitted into 802.11b DCF.

The number of packets transmitted by OAR in any transmission burst should be limited so
as to provide a fair channel access to all nodes.  OAR tries to achieve this by
allocating a fair temporal share of the channel. The fair temporal share is determined
as the maximum time the channel can be occupied if OAR were to transmit a single packet
at the base rate. The base rate of a channel is the lowest possible rate with which
data can be transmitted. The base rate of 802.11b channel is 2 Mbps. Thus the number of
packets sent in every burst is equal to the ratio of sending rate and the base rate.
Thus the burst is limited to at most 5 packets when the selected transmission rate is
11 Mbps. This guarantees that OAR inherits the same temporal fairness properties of
original 802.11 base protocol~\cite{arf}.

\figw{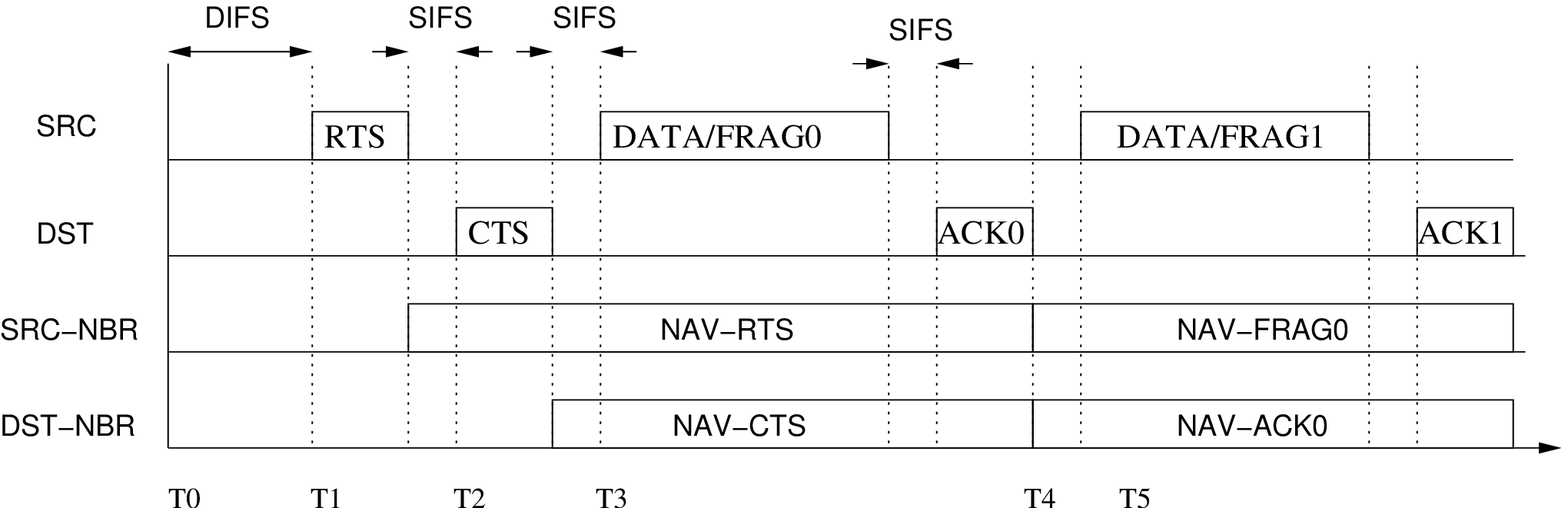}{6in}{\sl The fragmentation mechanism in 802.11b. Each fragment acts as a
virtual RTS for the next fragment and the ACK acts as a virtual CTS.}

In OAR, the transfer of a burst of packets is achieved by using the IEEE 802.11b
fragmentation mechanism. Figure~\ref{frag:fig} shows the timeline for transmission of a
fragmented data packet. Each fragment carries the duration information about the next
fragment to update the NAVs of neighboring nodes. The ACKs for fragments repeat the
same information for the benefit of nodes in the vicinity of receiver. Only the last
fragment and the last ACK do not carry this information. Also, each fragment except the
last one has {\em more fragments} flag in the frame control field set to 1 to indicate
the use of fragmentation mechanism. Thus, each fragment/ACK pair acts as a virtual
RTS/CTS.  Also the transmission gap between any two frames is SIFS. This ensures that
new transmission attempt by other nodes do not take place before all the fragments are
transmitted.

If the date rate is above the base rate, OAR sets the {\em more fragments} field of
each frame to 1 for an appropriate number of packets. Also, to disable defragmentation
at the receiver end {\em fragment number} field in each frame is set to 0. This enables
OAR to send a burst of packets without initiating contention. This provides the so
called {\em opportunistic contention gain}.  With performance gains because of rate
adaptation and opportunistic contention gains OAR outperforms ARF and RBAR. OAR
achieves throughput gains of around 40\% to 50\% over RBAR.

To realize these performance gains, it is necessary that the packets that are queued
with the network interface card are meant for the same destination node, otherwise,
packets destined for different nodes cannot be sent in a burst. The best place where
one can use OAR is the infrastructure wireless LAN where all packets are sent through
the base station. Since the destination of all upstream packets is a unique node, OAR
stands to have more performance gains as compared to the ad hoc wireless LANs. 802.11b
also support a 2-way handshaking data transfer which does not require RTS/CTS message
exchange. This mode is widely used when the packet size is smaller than some
configurable threshold.  On comparing with 2-way handshaking, OAR does not provide much
performance gains because OAR obtains performance gains by reducing the overhead posed
by RTS/CTS messages which do not get exchanged in 2-way handshaking.

\paragraph {\bf DCF+\\} The TCP {\em Self Collision} problem (mentioned earlier) occurs
because TCP ACK packets contend with data packets for the same channel in half duplex
mode of wireless LANs.  Haitao Wu et al.~\cite{haitao} propose a scheme called DCF+
which tries to provide a solution for this problem by modifying the MAC protocol.  
DCF+ tries to reduce the contention between data and TCP ACKs. In DCF, every node starts a
contention after sensing the channel to be idle for DIFS period. DCF+ aims at removing
the contention for TCP ACKs by giving them preferential treatment and hence avoid the
self collisions.

\figw{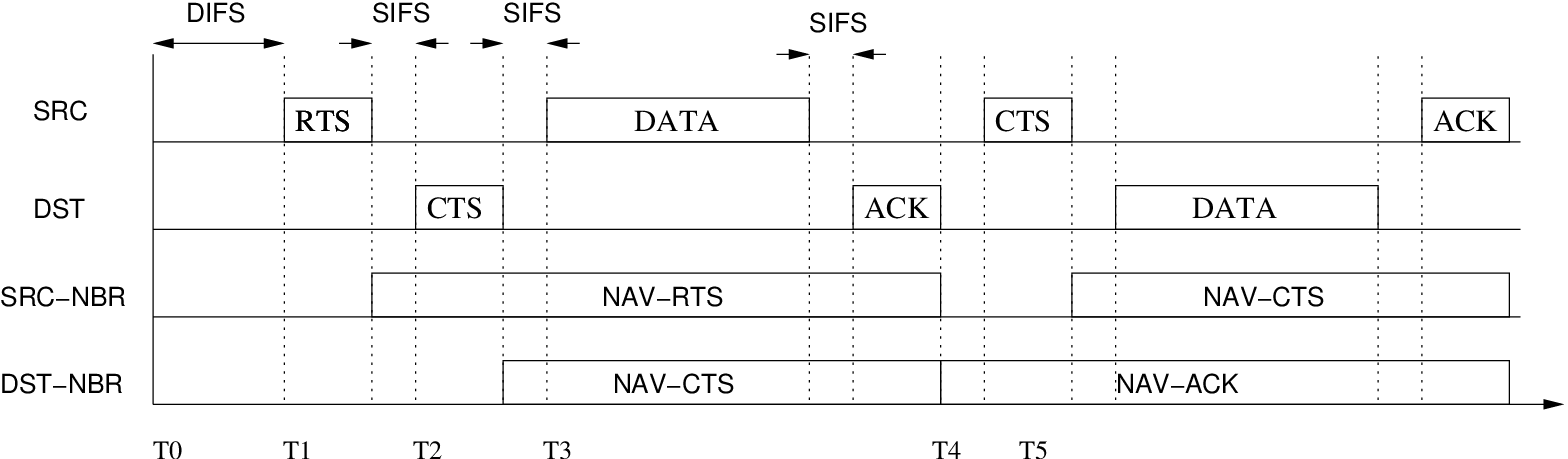}{6in}{\sl Frame exchanges in DCF+. ACK for data packet acts as an RTS for
TCP-ACK. This way TCP-ACKs do not need to contend for channel. The exchange terminates
with normal ACK for second transmission.}

Figure~\ref{dcfplus:fig} illustrates the packet exchanges in DCF+ protocol. Assume that
some node A sends a data packet (may be after RTS/CTS handshake) to some node B. Node
B, upon receiving this packet, needs to send back an ACK to node A. If the node B has a
packet ready in its queues for node A, it sends back an ACK with the duration field set
to some value which would set the NAVs for the nodes in its neighborhood. When such an
ACK arrives at node A, it responds back with a CTS so that the nodes in its
neighborhood set their NAVs. Upon receiving CTS, node B sends the data (TCP-ACK) to
node A which is acknowledged in normal way. This mechanism is backward compatible with
DCF as there are no new frames introduced. If a node does not understand DCF+, it
simply refrains from participating in DCF+. The ACK messages serve the purpose of RTS
messages for backward traffic, and thus, the contention is avoided for ACKs. Simulation
analysis for this protocol shows that it has around 5\% to 20\% performance gain over
conventional DCF. The performance gain increases with the increase in number of
wireless nodes. Also the DCF+ protocol is shown to be more fair compared to DCF.

This proposal requires the MAC layer to be aware of TCP in order to identify TCP-ACKs
and prioritize them. Moreover, in presence of fragmented frames this protocol cannot
function. Note that the use of duration field n ACKs by DCF+ and the fragmentation
mechanism will be conflicting with each other. The fragmentation mechanism in DCF uses
the duration field for the benefit of subsequent fragment transmission. It is not clear
what would be the behavior of DCF+ when fragments are encountered. This proposal
assumes that the probability of data and TCP-ACK collisions is very high in a
peer-to-peer TCP connection.  But in reality, the collisions are very limited because
of the binary exponential backoff.

\Ssse{Higher Layer Optimization} 

The default response of nodes to bit errors is to drop the erroneous frames if the
forward error correction cannot retrieve the original contents. But certain
applications like audio/video applications often prefer packets with partial errors
over lost packets. Performance of such applications can be greatly improved if the link
layer passes the corrupt packets up the protocol stack and leaves the decision about
whether to drop the packets or not to the applications. UDP~Lite~\cite{udplite} is a
protocol based on UDP which allows applications to specify the extent of data that
needs to be protected by checksum. This is achieved by providing the size count for
sensitive data in the header. For performance purpose, this sensitive data needs to be
at the beginning of the packet.  This way, an application can indicate its preference
about the treatment of checksum errors. By using UDP Lite like protocols applications
can still function while being resilient to the channel errors which do not affect
their correctness.

UDP lite works on the premise that most of the errors are created inside the node and
very few errors are created over the links. This is not true for wireless links.  To
function properly on wireless links, UDP Lite requires support from link layer drivers
either in the form of partial checksums or some mechanism to disable checksums.  To our
knowledge, there is no direct way of achieving this for 802.11b commercial cards. The
closest one can get to this is to enable the RF Monitoring mode on a card to receive
all packets. This is equivalent to sniffing functionality in wired networks.

Given the requirement, it is also feasible to incorporate the partial checksum
functionality in commercial wireless network interface cards. The MAC for wireless
networks can borrow the idea of partial checksumming from UDP-Lite and leave the
decision of dropping the frames to higher layers or to applications. This would greatly
benefit the media applications like Voice over IP or video streaming.

\Ssse{Infrastructure Solutions} 

The performance of wireless LANs can also be improved by careful structuring and
balancing the radio load. The maximum channel throughput in 802.11b LANs with multiple
nodes does not exceed more that 5 to 6 Mbps. A common technique to improve this
throughput is to aggregate multiple non-overlapping channels in same physical area. For
802.11b, atmost 3 non-overlapping channels can be aggregated to achieve a theoretical
performance of 33 Mbps. This technique is termed as {\em channel aggregation}.

In the {\em infrastructure (access-point)} wireless LANs all traffic needs to pass
through the access point. Thus, the performance of an infrastructure WLAN is limited by
the access point bandwidth. Usually, infrastructure mode wireless LANs are used as the
wireless access networks to the wired networks. A clear improvement is possible when
multiple access points are provided in a non-overlapped channel fashion. This is
similar to the technique used in cellular mobile systems. The throughput performance of
a cell can be improved by shrinking the cell size and increasing the cell
density~\cite{hpcn}.

Another scope of improvement in infrastructure WLANs is in reducing the load on the
access point by minimizing the intra-wireless network traffic. Each packet sent by a
wireless node to another wireless node consumes the bandwidth equivalent of two packet
transmissions, once upstream--to the access point, next downstream--from the access
point to the destination node. This is acceptable if the two communicating nodes are
mutually out of range. But if the nodes are in proximity, it a waste of scarce wireless
LAN bandwidth. In such scenarios, the nodes communicating with each other can establish
a transient ad hoc network for intra-wireless network traffic and remove the burden
from the access point. A. Raniwala et al.~\cite{raniwala} propose a scheme called
microAdhoc networking to achieve this. Some of the issues that need to be addressed in
this approach are; seeking the peer nodes for establishing an ad hoc network, channel
switching latencies, and maintaining the simultaneous membership of ad hoc as well as
infrastructure network.

\Se{Intelligent Collision Avoidance}

Typically wireless LAN Medium Access Control protocols are decentralized and
distributed protocols like 802.11b~\cite{80211b}, ALOHA~\cite{aloha}, MACA~\cite{maca},
MACAW~\cite{macaw}, etc. The primary objective of these protocols is to reduce the
collision that may occur when multiple nodes try to transmit data simultaneously. This
objective is achieved by coupling the carrier sense with {\em slotted randomizing} the
packet transmission instead of {\em exclusive medium access} semantics to trade the
complexities and overheads of such mechanism with a low probability of collisions.

All distributed wireless MAC protocols encounter significant problems like {\em hidden
node problem} and {\em exposed node problem}~\cite{maca} because of this distributed
and randomized nature. Hidden nodes result in an improper carrier sense and hence
potential collisions, whereas, Exposed nodes result in a misdirected carrier sense and
hence unwarranted refraining from transmissions. Both scenarios result in reduction of
the overall throughput of the wireless channel.

The Collision Avoidance~\cite{maca} extension to CSMA protocols provides an elegant
solution to the hidden node problem. Collisions are avoided by exchanging additional
management messages like Request To Send:RTS and Clear To Send:CTS dueling as the
channel reservation messages directed to the neighboring nodes. On hearing these
RTS/CTS messages the neighboring nodes, which could potentially be the hidden nodes,
refrain from any transmission activity for a period that is mentioned in the RTS/CTS
messages. A pictorial representation of collision avoidance in shown in Figure
\ref{access:fig}. This collision avoidance mechanism, although very elegant and simple,
comes with an overhead of extra messages which is not justifiable when the required
data transmission duration is small.  Usually network interface equipments implementing
CSMA protocols follow a threshold size for data transmission sizes above which CA
(virtual carrier sensing) is enabled and otherwise it is turned off.

\figw{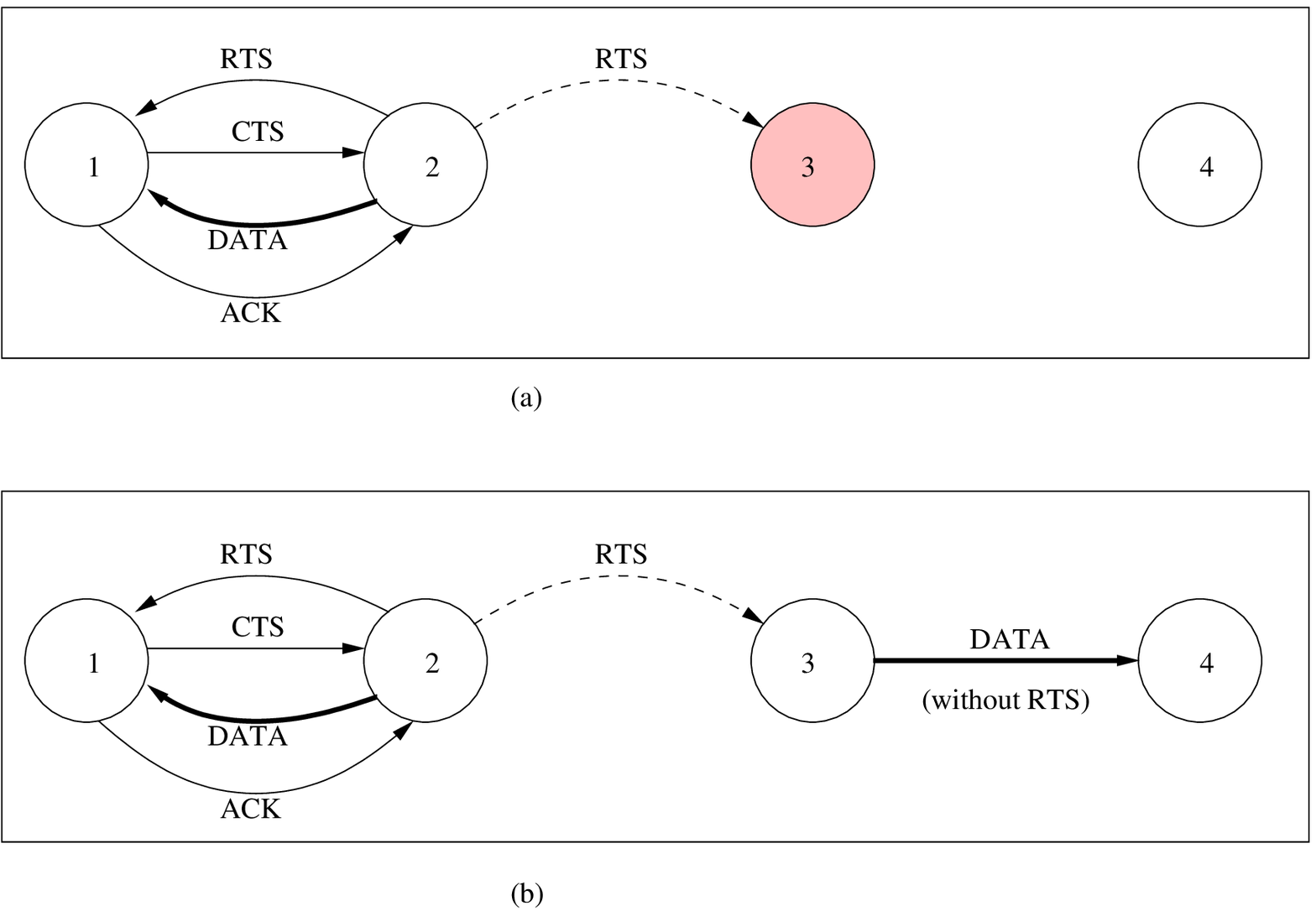}{6in}{\sl (a) Basic channel access mechanism in CSMA/CA. Node 2 broadcasts
an RTS message which is received by 1 and 3. The intended recipient node 1 responds
with CTS. Node 2 then initiates a transmission to node 1. Finally node 1 can send back
an ACK message for additional reliability. All the while, Node 3 holds back its
transmission to node 4 even if it would not have caused any problems. (b) In the
improved channel access mechanism, node 3 deduces the directionality of transmission of
Node 2 depending on the inability of CTS from 1 to reach node 3. Enabling a parallel
data transmission to node 4.}

Exposed node problem, on the other hand, occurs because of a misdirected carrier sense
which results in refraining from data transmission even in the situations where
collisions would not occur causing a reduced overall channel utilization. To the best
of our knowledge, there exists no practical solution to this problem. Further the
collision avoidance mechanism, if used, results in a conservative behavior by all the
nodes that are present in the vicinity of the nodes that are currently using the
wireless channel. In this work, we propose a mechanism called the {\em Intelligent
Collision Avoidance} (ICA) which tries to improve the channel utilization by allowing
the exposed nodes to carry on data transmissions in parallel whenever possible. This
mechanism leverages on the RTS/CTS messages exchanged during the collision avoidance
phase to deduce the directionality of transmission and RTS/CTS threshold values to
carry out parallel data transmissions.


\Sse{Proposed Scheme}

The root cause of exposed node problem is the misdirected carrier sense by nodes
causing unwarranted refraining from data transmissions. In the pure CSMA MAC, it
becomes impossible to deduce the location of intended receivers of any data
transmission. But in CSMA/CA protocols it is possible to decipher the directionality of
the data transmissions by leveraging on the control messages exchanged during the
collision avoidance phase. We assume that the wireless transmission characteristics are
symmetric, i.e. the mutual signal levels that are observed at any two nodes are same
provided the transmission power used by both the nodes is same.

In our scheme we intend to rely on the signal strength of the CTS message that is sent
by the receiver node to decipher the directionality of data transmissions and hence
asses the impact of any parallel data transmission on the reception of receiver node.
In basic CSMA/CA protocol whenever a node hears an RTS message it refrains from
transmitting any data. But instead it can rely on the absence of CTS response to
decipher that the receiver happens to be out of range for transmissions from this node
and hence it is an exposed node for this particular instance of transmission. In such
case the exposed node can initiate a parallel data transmission improving the channel
utilization. In Figure \ref{access:fig} node 3 is the exposed node, node 1 is the
receiver, and node 2 is the sender.

Although the exposed node can safely transmit data, it can not receive any data because
of other ongoing data transmission. Thus the exposed node is accompanied by a hidden node
which has already initiated data transfer. In Figure \ref{access:fig} node 3 is the
exposed node and node 2 is the hidden node for node 4. Thus, any transmission by
node 4 intended for node 3 will not be heard by it because of ongoing transmission from
node 2. This would hamper tranmissions like CTS from node 4.
This forces node 3 to resort to data transmission without using collision avoidance. 
Further, if the MAC protocol uses link level ACKs then the exposed node can turn into a 
hidden node for the ACK traffic.
This problem can be resolved by limiting the data transmission by the exposed node so
that the ACK is not garbled. This requires a proper estimation of the data transmission
time and the time at which the ACK will be transmitted.

The timing estimate can be easily obtained from the RTS frame that is initially
transmitted by the sender node.  The exposed node can now limit the data transmission
for this duration. The limiting size can be determined by using the existing
fragmentation logic in the MAC.

The medium access algorithm for every node intending to transmit data in this case is:
\begin{itemize}
\item Sense the channel for the  DIFS period
\item If an RTS message is seen, wait for CTS timeout, else follow the usual DCF procedure.
\item In the absence of CTS message, deduce that the node is an exposed node
and start data transmission. Fragment the data so that the data transfer ends
exactly when the data transmission from the node sending RTS also ends. 
\item Wait for ACK. If an ACK is not received assume that the data did not get transmitted
properly else follow the same procedure for remaining portion of data.
\end{itemize}

\figw{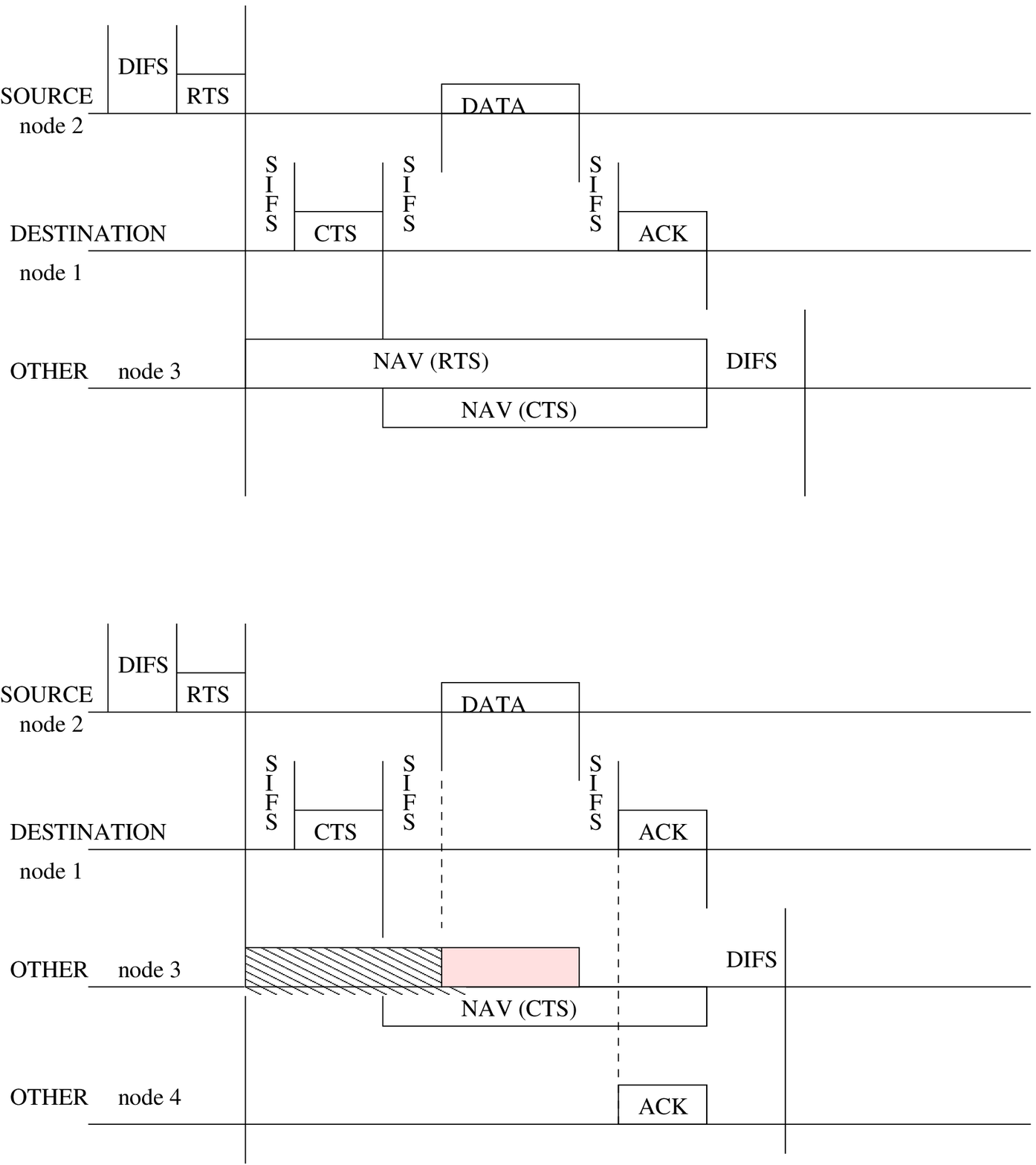}{6in}{\sl Timing diagram for basic and intelligent CSMA/CA mechanism. Node
3 is the exposed node trying to carry out a parallel data transmission with Node 2.}

The resulting timing diagram is depicted in Figure \ref{timing:fig}.

\Sse{Performance Simulation}

We are in the process of analyzing this modification to protocol. We have modified the
NS-2 simulator with an extension for ICA.  We have observed a two-fold performance
gains for simple string topology. We plan to evaluate our scheme for the following
aspects:

\begin{itemize}
\item{Scalability: The extent to which parallel data transmissions are possible. How
different topologies and different number of nodes involved affect our scheme.}
\item{Range: One of issues here is the impact of ratio of hearing range and sense range.
If the sense range of wireless NICs is much higher than the hearing range then it is not
clear how parallel transmissions are going to affect each other. We are relying on
the radio capture effect to decipher packet transmissions correctly in the event parallel
transmissions interfere with each other.}
\item{Mobility: The impact of mobility is unclear.  We are relying on long coherence
period to alleviate the impact of mobility on the overall scheme.}
\item{Power consumption: The Intelligent Collision Avoidance scheme promises throughput
gains at an expense of additional power. We intend to analyze the exact impact of the schema 
on the overall power consumption in the network.}
\end{itemize}

\Se{Conclusions}

Though IEEE 802.11b is the most dominant wireless LAN technology today, it is evident
that is not a well matured technology and has a fair share of its own problems.

In this survey we focused on certain aspects of 802.11b like performance issues, QoS,
and fairness.  We highlighted the main issues in these areas and compared the current
research aimed at performance enhancement, QoS provisioning, and fair scheduling. Most
of the solutions we reviewed are closely related to the MAC layer enhancements. It is
possible to solve all these solution by enhancing/modifying different aspects of the
protocol stack. As an example we cite various TCP performance related solutions at
different levels. It is our observation that the solutions are effective and simple if
these are provided at the source of the problem itself.  The basic distinguishing
factor for wireless media from wired media is the nature of the medium itself. Thus, it
is our belief that most of the solutions for the problems introduced because of
quirkiness of wireless medium should be provided closest to the medium and that is the
MAC layer of the protocol stack.

We also suggested a scheme (ICA) to improve the overall network throughput by slight enhancements to the MAC layer. ICA is inspired by other MAC layer schemes like RBAR, OAR, and MACAW protocol.  In future, we plan to evaluate ICA to verify its effectiveness. We also plan to study
various scalability, mobility, and other related issues.

\bibliography{wlanrpe}
\bibliographystyle{unsrt}

\end{document}